\documentclass[12pt,preprint]{aastex}

\shorttitle{Cepheid Metallicity Dependence}
\shortauthors{Mager, Madore \& Freedman}

\begin{document}

\title{The Metallicity Dependence of the Cepheid P-L Relation in M101}

\author{Violet A. Mager\altaffilmark{1,2,3} \email{mager@susqu.edu}}
\author{Barry F. Madore\altaffilmark{1} \email{barry@obs.carnegiescience.edu}}
\and
\author{Wendy L. Freedman\altaffilmark{1} \email{wendy@obs.carnegiescience.edu}}

\altaffiltext{1}{Carnegie Observatories, 813 Santa Barbara St., Pasadena,
CA 91101}
\altaffiltext{2}{Ohio University, Department of Physics and Astronomy,
Athens, Ohio 45701}
\altaffiltext{3}{Susquehanna University, Department of Physics, 514
University Avenue, Selinsgrove, PA 17870}

\begin{abstract}
The impact of metallicity on the
Cepheid Period-Luminosity (P-L) relation is investigated using HST ACS V and
I images of M101. Variations in the reddening-free Wesenheit parameter (W),
which is employed as a proxy for luminosity, are examined as a function of the
radial distance from the center of M101 (and thus metallicity). We
determine that there is no dependence of the slope on metallicity. However, 
the intercept is found to depend on metallicity by 
$\gamma_{VI} = -0.33 \pm 0.12$ mag dex$^{-1}$ and 
$\gamma_{VI} = -0.71 \pm 0.17$ mag dex$^{-1}$ using 2 and 3 sigma rejection 
criteria,
respectively.  Sigma-clipping impacts the derived metallicity dependence, and
the 2-sigma criterion applied likely mitigates blending, particularly in the
crowded inner regions of M101. A metallicity-corrected distance for M101
is obtained from 619 Cepheids ($\mu = 28.96 \pm 0.11$), a result that 
agrees with
the recently determined SN Ia distance. The metallicity effects described can
be bypassed by working at near and mid-infrared wavelengths (e.g., the
Carnegie Hubble Program).
\end{abstract}


\section{Introduction}

Determining accurate distances to astronomical objects is a fundamental but 
challenging problem in astronomy, having far-reaching 
consequences not only for our understanding of individual objects, but 
more generally for the fundamental parameters of the Universe such as the 
Hubble Constant. The Cepheid period-luminosity (P-L) 
relationship has been used extensively in calibrating the extragalactic 
distance scale (e.g., Freedman \& Madore 2011). 
There is, however, controversy over whether the slope and absolute 
intercept of this relation are universal, and to what extent they vary
if they are not (e.g. Sandage et al. 2009, 
Tammann et al. 2008, and the references therein). We address this
important question in this paper. 

Theoretical models suggest that differences in the Cepheid P-L relation 
slope and zero-point can potentially be caused by variations in chemical 
composition (e.g. Marconi 2005, Caputo 2008), which could be quantifiable
by investigating how these quantities depend on host-galaxy metallicity.
However, more recent results from these groups (Bono et al. 2010), and others 
(e.g., Majaess et al. 2011, Pellerin \& Macri 2011a; 2011b, 
Udalski et al. 2001) argue that there is an insignificant correction for 
metallicity in VI, and that claims to the contrary may be due to crowding. 
This is in agreement with the result of Rizzi et al. (2007), who find
that tip of the red giant branch (TRGB) distances agree well with Cepheid 
distances when a metallicity
correction is not applied to the Cepheid P-L relation.
Many other studies find no correction on the P-L slope, but a 
significant correction in intercept typically around -0.25 mag dex$^{-1}$ 
(Freedman \& Madore 2011, 
and the references therein), although the results from some authors
differ from this considerably. For instance, Romaniello et al. (2008) 
find the opposite sign on the correction in V (with higher metallicity stars 
appearing fainter), and Shappee \& Stanek (2011), find a relatively steep 
correction in M101 of -0.80 or -0.72 mag dex$^{-1}$ using two different 
methods. Gerke et al. (2011) also find a steep dependence on metallicity
of -0.56 mag dex$^{-1}$ in M81. Based on these contradictory results, it
is apparent that this issue is still controversial, and 
requires further investigation.

The Hubble Space Telescope (HST) Key Project (Freedman, et al. 2001) 
sought to improve the extragalactic distance scale using Cepheids discovered 
in HST Wide Field and Planetary Camera 2 (WFPC2) data to calibrate several 
secondary distance measurement methods. They list uncertainty in the 
metallicity calibration of the Cepheid period-luminosity relation as 
one of their largest sources of error. 
Following Freedman \& Madore's (1990) work on M31, the HST Key Project 
in part undertook measurements of the metallicity dependence of the P-L 
relation using HST WFPC2 observations of M101 in V (F55W) and I (F814W) 
(Kennicutt et al. 1998). They compared the P-L relations in two radially 
separated fields of differing metallicities, finding a dependence of the 
zero-point on metallicity of $-0.24 \pm 0.16$ mag dex$^{-1}$. This study 
involved relatively small number statistics (26 Cepheids in the outer field, 
and 50 in the inner field), and the final result relied on data in only
two radial zones (i.e., the difference between the outer and inner fields). 
The statistical significance of this and other metallicity tests at the
time were low, so Freedman et al. (2001) adopted an overall metallicity 
correction (with a very large uncertainty) equal to the midrange of all the 
published empirical values ($-0.2 \pm 0.2$ mag dex$^{-1}$), with the
caveat that further investigation into the metallicity dependence 
was needed. Additionally, Macri et al. (2001) suggested that blending
could be an issue with the M101 images. They tested this by creating
artificial images of their own WFPC2 data of M31 and M81, as it would appear 
at the same distance and exposure time as the M101 data. They found a
difference in the P-L zero-point between the original and artificial images 
that could account for a significant fraction of the metallicity dependence
calculated by Kennicutt et al. (1998).

These issues with the WFPC2 M101 images motivated the acquisition of the 
data used in this project: two fields in M101 imaged by HST in the same 
filters as with Freedman et al. (2001), but with the superior imaging 
quality of the Advanced Camera for Surveys (ACS). The higher resolution and 
larger number statistics achievable from this 
data set provide the opportunity to conduct a more refined analysis of the 
metallicity dependence of the P-L relation over a variety of metallicities 
available across the field of views.

\section{Data}

Two separate fields in M101 were observed with the HST/ACS Wide Field
Camera (WFC). Each
field was visited on 12 occasions ranging from $\sim 1 - 5$ days apart. 
On each visit (epoch), two cosmic-ray-split exposures were taken in 
both F814W and F555W, with a total exposure time per visit of 724 sec 
in F814W and 1330 sec in F555W. Figure 1 shows the location of these ACS 
fields overlaid on a DSS\footnote{The Digitized Sky Surveys were produced at
the Space Telescope Science Institute under U.S. Government grant NAG W-2166.
The images of these surveys are based on photographic data obtained using
the Oschin Schmidt Telescope on Palomar Mountain and the UK Schmidt
Telescope. The plates were processed into the present compressed digital
form with the permission of these institutions.} image of M101.

We first masked the bad pixels in the calibrated and flat-fielded 
individual images from the HST pipeline. 
Then, for each epoch, the two images in each filter were integer-shifted and 
combined using IRAF.\footnote{IRAF is distributed by the National
Optical Astronomy Observatories, which are operated by the Association
of Universities for Research in Astronomy, Inc., under cooperative
agreement with the National Science Foundation.} We registered the images
for each field to within 1 pixel, and used DOLPHOT\footnote{By Andrew 
Dolphin, \url{americano.dolphinsim.com/dolphot}} to identify stars
and obtain their Johnson-Cousins V and I magnitudes at each epoch. 
For this process we used the recommended settings given in the 
DOLPHOT/ACS User's 
Guide\footnote{\url{americano.dolphinsim.com/dolphot/dolphotACS.pdf}}.

In order to avoid including objects that were too elongated, 
sharp, or extended to be stars (such as cosmic rays or galaxies), we
selected only those detections that were determined by DOLPHOT to
be a good star (type = 1). We further eliminated highly uncertain magnitudes 
from our analysis by excluding any individual images that were assigned a
flag of 8 or greater by DOLPHOT, which refers to extreme cases of
problems affecting the photometry (such as too many bad or saturated pixels). 
Any stars that had fewer than two good individual magnitude measurements were
eliminated altogether. After applying these selection criteria, there
were 425,803 objects detected in Field 1, and 487,238 in Field 2.

\subsection{Selecting Cepheid Candidates}

We selected possible variable stars using the variability index (L)
proposed by Stetson (1996), which combines the kurtosis (K) and the Stetson 
variability index (J) into a single, more robust measurement of a 
star's variability (Stetson 1996, Welch \& Stetson 1993). The J term 
determines variability likelihood based on how much the individual magnitudes
vary from the average magnitude, and whether the form of the variations is
correlated between pairs of observations or filters. The kurtosis encapsulates 
the overall shape of the magnitude distribution, which will contribute
to higher L values if the light curve is shaped similar to a square-wave or
sinusoid, and lower values if it is composed of  purely random (Gaussian) 
variations 
or only a single magnitude spike. The variability index L, is thus:
\begin{equation}
L = \frac{J \times K} {0.798} \frac{\sum_{k=1}^{N_p} w_k}{w_{max}},
\end{equation}
\begin{equation}
K = \frac{(1/N) \sum_{i=1}^N |\delta_i|}{\sqrt{(1/N) \sum_{i=1}^N \delta_i^2}}, 
\end{equation}
\begin{equation}
J = \frac{\sum_{k=1}^{N_p} (w_k ~ sgn(P_k) \sqrt{|P_k|})}
{\sum_{k=1}^{N_p}  w_k}.
\end{equation}
The sums are either taken over each of the ith observations regardless of 
filter, or over each of the kth pairing of observations, which in our case
were comprised of the observations in V and I that were taken in each epoch. 
N is the total number of good observations in both filters, and N$_p$ 
is the total number of pairs that contained at least one good observation 
in one filter.

The residual of the ith magnitude measurement (m$_i$) from the 
weighted average magnitude in that filter ($<m>_x$) is given by:
\begin{equation}
\delta_i=\sqrt{\frac{N_x}{N_x-1}} ~ \frac{m_i - <m>_x}{\sigma_i},
\end{equation}
where N$_x$ is the total number of observations in the xth filter that
contributed to the mean, and $\sigma_i$ is the photometric error on m$_i$.

The average magnitude was calculated, as suggested by Stetson (1996), 
through an iterative process of weighting each magnitude measurement by  
(1+($\delta_i$/2)$^2$)$^{-1}$, which applies less weight to outliers
from the mean. Each
magnitude was also multiplied by $\sigma_i^{-2}$ in order to apply less 
weight to more photometrically uncertain measurements. We found 
that the average calculations converged to within 0.001 mag in $\lesssim 6$ 
iterations.

The quantity P$_k$ combines the individual residuals within each pair of 
observations. In the case where the kth epoch contains good magnitude 
measurements in both filters (m$_i$ and m$_j$), it is given by:
\begin{equation}
P_k=\delta_i \delta_j
\end{equation} 
If the kth epoch only contained one good magnitude (m$_i$) in a single
filter, it is instead given by:
\begin{equation}
P_k=\delta_i^2 - 1 
\end{equation}

Furthermore, the variability index is weighted by w$_k$, where, $w_k=1$ if 
there are good measurements in both filters, $w_k=0.5$ if there is only one 
good measurement, and $w_k=0$ if there are none. This is normalized by 
w$_{max}$, which is the total weight a star would have if it had good 
magnitude measurements in all of the available images (in our case, this
is equal to the number of epochs, or 12).

We chose a cutoff of L$ > 1.2$ for the variable candidates, which corresponds
to the range of values for our particular dataset within which L is 
correlated with the standard deviation, and is thus less likely to be 
dominated by random noise. We applied further constraints on the
sharpness of the star as measured by DOLPHOT in V (accepting values of 
--0.3 to 0.3), and the crowding as measured in both V ($<0.85$) and 
I ($<1.00$). Objects with sharpness and crowding measurements outside of these
ranges were empirically found to contain a large number of extreme 
outliers from the expected parameters for Cepheids. Also, in order to 
reject stars that are not within the instability strip, and thus are likely 
not Cepheids, we then applied cutoffs as determined from the star's
position on the color-magnitude diagram: $0.5\leqq$(V--I)$\leqq 1.25$,
and m$_I \leqq 25.3$ (see Figure 2). The application of these constraints 
results in 650 candidate Cepheids in Field 1, and 975 in 
Field 2. We further constrained the sample size through visual inspection of
the data, as explained in Section 2.2.

\subsection{Measuring Periods and Magnitudes}

In order to find trial periods for each of the candidate Cepheids, we
employed the light curve string-length minimization technique originally
introduced by Lafler \& Kinman (1965). This method works on the principle that
magnitude data points ordered by phase for a chosen period will have a
minimum total path length between them when the chosen period is equal 
to the true period. Stetson (1996) improved upon the original formula
for the string length by making it more robust against uncertain or corrupt
data points, as well as to data that are not evenly sampled in
phase space. After ordering the data points in ascending phase for a
chosen trial period (P), this string length within a filter x is calculated 
using:
\begin{equation}
S(P)=\frac{\sum_{i=1}^{N_x-1} w(i,i+1) |m_i-m_{i+1}|}{
\sum_{i=1}^{N_x-1} w(i,i+1)},
\end{equation}
where w(i, i+1) is a weight factor given by:
\begin{equation}
w(i,i+1)=\frac{1}{\sigma_i^2+\sigma_{i+1}^2} \times 
\frac{1}{\phi_{i+1}-\phi_i+N_x^{-1}}.
\end{equation}
The phase of the ith observation is $\phi_i$ = ($t_i - t_1$)/P, where
$t_i$ is the observation's Julian date at the mid-point of the exposure.

We calculated string lengths in V for each Cepheid candidate using trial 
periods between 2 and 30 days, in steps of 0.01 days. This period 
range is limited by the sample itself, within which shorter period (fainter) 
Cepheids are lost in the noise, and longer periods are unobtainable because 
they extend beyond the window of observations. Figure 3 shows an example
of the string length as a function of trial period, and Figure 4 shows
the light curve and phase diagram for this star using the period obtained
through string length minimization. The true period 
should in principle correspond to the absolute minimum value of S(P), but in
practice this is not always the case: rather, the true period could correspond
to any of the lowest local minima. Given this, we visually reviewed the
phase curves for periods corresponding to the first few local minima,
choosing the one that produced the smoothest, most ordered phase curve. 

At this point we eliminated more stars from our Cepheid candidate list, 
setting aside those that had periods greater
than the window of observations, as well as those that had 
periods $\lesssim 4$ days, which would only introduce errors into our 
final analysis as their light curves are extremely noisy. We also rejected
variable stars that had profile shapes that were
clearly not those of Cepheids, such as profiles with a very slow rise
in brightness and a very fast decline, or profiles with a long flat peak
or a single brief increase (or decrease) in brightness within an otherwise
flat profile. Also, extremely uncertain data were rejected by eliminating 
stars with measurements that were too noisy or had photometric 
errors that were too large to be certain of the correct period. We also
rejected stars where the V and I light curves were uncorrelated over a 
significant
portion of the curve. After applying these additional constraints, we
obtained a final list of 292 Cepheids in Field 1, and
327 Cepheids in Field 2. The physical locations of these Cepheids are
shown in Figure 1 (small circles), and their locations on the color-magnitude
diagrams are displayed in Figure 2 (blue filled circles). 

Average magnitudes for these Cepheids were determined using GLOESS
(e.g., Persson et al. 2004), an algorithm
that smoothly fits a curve to the data points using local regression, 
and then determines the
average magnitude of the fit. Examples of some of the phase-ordered 
Cepheid light curves and their fits are given in Figure 5. All of the
light curves are available in the online edition. Table 1 lists
the Celestial coordinates, average magnitudes, periods, V and I
amplitudes, radial distance from the center of the galaxy, and
log(O/H) metallicity for ten
of the Cepheid candidates (full table available online).

We check our magnitudes and periods for accuracy by comparing our
results to Stetson et al. (1998), whose HST WFPC2 images overlap
with our Field 1. Figure 6 shows a comparison of the periods and V and I
band magnitudes of the Cepheid stars with positions in the registered 
images that match within 7 pixels. We find no systematic difference between
the two samples, with a median residual in the periods of only 0.07 days, and
a standard deviation from zero of the difference between the samples
of 0.13 mag, 0.24 mag, and 0.74 days for V, I, and period, 
respectively. 

\section{Results}

\subsection{Metallicity Dependence of the Wesenheit P-L Relation}

We use the Wesenheit parameter (Madore 1982) for luminosity in our 
P-L relations, which is unaffected by reddening, and thus results in 
lower scatter in our plots than V or I. Use of the Wesenheit parameter
has also been shown to produce a more linear P-L relation than that obtained
from the individual pass-bands (Ngeow \& Kanbur, 2005), and a constant
slope regardless of metallicity (Bono et al. 2010). It is particularly
robust in V and I: Bono et al. (2010), for instance, find a metallicity
gradient of the zero-point in W$_{VI}$ of only $-0.03 \pm 0.07$ mag dex$^{-1}$, 
compared to their more significant gradient in W$_{BV}$ of $-0.52 \pm 0.09$ 
mag dex$^{-1}$. In light of its robustness relative to straight P-L relations, 
Bono et al. (2010) suggest that the Wesenheit parameter provides a 
preferable method of measuring the distance modulus of galaxies relative 
to the LMC, as long as 
any metallicity dependence of the zero-point is taken account.
The Wesenheit parameter is calculated for each star using their V and I 
magnitudes with:

\begin{equation}
W = V - {A(V)\over E(V-I)} (V-I)
\end{equation} 

A global total-to-selective absorption ratio A(V)/E(V-I) of 2.45
was used for both the Key Project (Freedman, et al. 2001) and
the independent measurement of the metallicity dependence in M101
conducted by Shappee \& Stanek (2011) on the same dataset used in
this paper. We adopt
the same value for consistency.

Figure 7 shows W as a function of the log of the periods for all of our 
Cepheid candidates, including a linear least-squares fit to the data that 
iteratively excludes all data-points with greater than 2-sigma 
deviation from the fit. This results in an overall P-L relation of:

\begin{equation}
W = -3.193 (\pm 0.025) (log(P) - 1) + 23.152 (\pm 0.006)
\end{equation}

 
For comparison, we list the slopes for Wesenheit VI P-L relations derived from 
different galaxies in Table 2. The relation for non-Lutz-Kelker bias corrected 
Galactic Cepheids was obtained by Benedict et al. (2007), and we derived
W for the LMC and SMC by applying Equation 9 to the V and I P-L relations 
determined from OGLE III data by Soszynksi et al. (2008) and (2010),
respectively. Our slope for M101 of $-3.193 \pm 0.025$ falls within the
mutual 1-$\sigma$ uncertainties of the Galactic and LMC slopes, and is
just outside of 1-$\sigma$ for the SMC. The SMC has a particularly
low metallicity, but there is no apparent trend of slope with the average 
iron metallicity of 
each galaxy, as given in Table 2 and measured by Mottini et al. (2006). 

Even when using the reddening-free Wesenheit parameter, there is
significant scatter in the P-L plot in Figure 7. Some of this may be due to 
differences in metallicity, which depends on the location of the star 
within its host galaxy, where metallicity in general increases toward 
the galaxy center. We investigate these potential metallicity effects by 
separating the Cepheids into $\sim 1\arcmin$-wide annuli according to 
their radial distance 
from the center of the galaxy, and then recalculating the P-L relation 
for each of these distance bins. The average [O/H] metallicity of each star is
calculated using the relationship derived by Kennicutt et al. (2003)
from abundance measurements of HII regions in M101:

\begin{equation}
log(O/H) = 8.76(\pm 0.06) - 0.90(\pm 0.08)(R/R_o)-12
\end{equation}  

We use a disk scale length, R$_o$, of 14.85\arcmin, or 32.4 kpc at 
Kennicutt et al.'s (2003) adopted distance of 7.5 Mpc to M101. 

The slope of the P-L relations is plotted in Figure 8 for the 
$\sim 1$ arcminute annuli with blue filled circles, as a function
of the average log(O/H) metallicity of each of the Cepheids in each
distance bin. 
To determine if the dependence of the slope on metallicity varies with
bin selection, we also plot the slopes for four other sample sets
chosen in different ways: 4 annuli with $\sim 124$ stars in each (black 
open squares), 6 annuli with $\sim 83$ stars in each (green open circles), 
8 annuli with $\sim 64$ stars in each (red open triangles), and 10 annuli 
with $\sim 52$ stars in each (magenta crosses). We see no difference in
behavior of the slope vs. metallicity with distance bin selection, and
no significant dependence of the slope in any of the sample sets on 
metallicity. As such we adopt the slope calculated from the entire sample
(from Eq. 10) for each of the $\sim 1$ arcminute-sized distance bins. 
This is consistent with the lack of a trend in slope as a function of 
average galaxy metallicity as shown in Table 2. 

Figures 9 through 12 demonstrate the P-L relation for each of the
distance bins, with the slope of the fit fixed to that from the entire 
dataset. The fits deviate from that of the entire dataset (thick
long-dashed line) in a systematic way, with the intercept becoming 
progressively fainter as the distance from the center of the galaxy increases. 
It is also of note that there is a smaller proportion of long-period to 
short-period Cepheids at progressively larger radial distances. This
demonstrates the radial composition gradient within the galaxy, as higher 
metal abundances have been shown to produce a higher frequency of brighter, 
longer period Cepheids (Becker et al. 1977).

The intercept of the P-L relations is plotted in Figure 13 as a function
of the average log(O/H) metallicity in each distance bin. Here we do
find a significant dependence on metallicity, with a weighted 
linear least-squares fit to the data yielding a slope of 
$\gamma_{VI} = -0.33 \pm 0.12$ mag dex$^{-1}$.
This corresponds to an increase in brightness of Cepheids with increasing
metallicity, and decreasing distance from the galaxy center.

The sample choice above relies on only four distance bins, with
a variable number of Cepheids in each. To determine if this has an
effect on our results, we
calculate the P-L relation intercept as a function of metallicity 
for several different samples of Cepheids chosen in different ways, using
the same distance bins as those selected for Figure 8.
The results are plotted in Figure 14, and include the above 
$\sim 1$ arcminute annuli (blue filled circles) that have a variable number 
of stars in each (49 to 204), and four other sample sets with variable annulus
sizes selected to have roughly the same number statistics in each bin:
4 annuli with $\sim 127$ stars in each (black open squares), 6 annuli 
with $\sim 85$ stars in each (green open circles), 8 annuli 
with $\sim 64$ stars in each (red open triangles), and 10 annuli with
$\sim 52$ stars in each (magenta crosses). Weighted linear least-squares
fits were determined for each sample, and plotted in the figure with the
same color as their corresponding data points. The slope varies between
the samples from -0.30 to -0.42 mag dex$^{-1}$, which agrees within 
the uncertainties with the 1 arcminute bin size result of
$\gamma_{VI} = -0.33 \pm 0.17$ mag dex$^{-1}$. As such, we conclude that 
there is no significant dependence on sample selection. 

\subsection{Distance to M101}

We find the distance to M101 with respect to the LMC using the
LMC P-L relations below (V$_o$ and I$_o$), as determined by Soszynski 
et al. (2008) from 
the 3361 Cepheids in OGLE-III (Optical Gravitational Lensing Experiment Survey).

\begin{equation}
V_o = -2.762 (\pm 0.022) log(P) + 17.530 (\pm 0.015)
\end{equation}
\begin{equation}
I_o = -2.959 (\pm 0.016) log(P) + 16.879 (\pm 0.010)
\end{equation}

Applying Equation 9 to these V and I P-L relations, with an adopted
A(V)/E(V-I) of 2.45, yields a reddening-free Wesenheit P-L relation 
for the LMC of

\begin{equation}
W_o = -3.245 (\pm 0.035) (log(P) - 1) + 12.690 (\pm 0.042).
\end{equation}

If we adopt a true distance modulus to the LMC of 18.48 mag
(Freedman et al. 2012, Monson et al. 2012), as determined
from the Cepheid P-L relation in the infrared with the Spitzer Space 
Telescope (Scowcroft et al. 2011), then the 
non-metallicity corrected distance modulus to M101 is

\begin{equation}
\mu = W + 3.245 (\pm 0.035) (log(P) - 1) + 5.790 (\pm 0.042).
\end{equation}

We calculate this distance modulus for each Cepheid in our sample,
and plot the results as a function of metallicity in Figure 15. An iterative
linear least-squares fit to the data, with a 2-sigma rejection level, 
is shown in the figure. There is a small, but significant dependence of the
derived distance modulus with metallicity, corresponding to larger
moduli at lower metallicities, or greater radial distances from the
galaxy center. We therefore define our distance modulus to M101 as that
derived from the fit to the data at the metallicity of the LMC. As in
the Key Project (Freedman et al. 2001), we adopt a metallicity of 
log(O/H) + 12 = $8.50 \pm 0.15$ for the LMC. At this metallicity,
the distance modulus for M101 is 
$\mu = 28.96 \pm 0.11$. 
This is within the range of values from the 
Supernova Ia distance of 28.86 to 29.17 mag (Matheson et al. 2012), and
agrees with the mean distance modulus of $29.18 \pm 0.31$ calculated by
the NASA/IPAC Extragalactic Database (NED) from 46 independent distance 
measurements. It is, however, outside the mutual errors of 
the TRGB distance of $29.30 \pm 0.01$ (random) $\pm 0.12$ (systematic) 
obtained by Lee \& Jang (2012).



\section{Discussion of Results}

We find no significant dependence of the slope of the Wesenheit P-L
relation on metallicity, which is consistent with Bono, et al.'s (2010)
findings from the combination of 87 independent Cepheid data sets.
Theoretical models predict a steepening of the P-L relation with
increased metallicity (e.g. Caputo 2008), but the use of the Wesenheit
parameter largely removes these effects (Madore \& Freedman 2009,
Bono et al. 2010). 
Shappee \& Stanek (2011) do find a modest increase in slope in the 
Wesenheit P-L relation of M101 with increasing metallicity of
$3.0^{+1.7}_{-1.8}$ mag log(day)$^{-1}$, but mention 
that the result is tenuous enough that it could be due to a statistical 
fluctuation in the data. As such, we consider our results to be consistent.

We find a significant dependence of the intercept of the Wesenheit
P-L relation on metallicity of $\gamma_{VI} = -0.33 \pm 0.12$ mag dex$^{-1}$, 
which is
steeper than, but in agreement with, the value of $-0.24 \pm 0.16$ mag
dex$^{-1}$ found through the HST Key Project (Kennicutt et al. 1998).
This is a weaker dependence than that obtained by Shappee \& Stanek's (2011) 
from the same dataset ($-0.80 \pm 0.21$ (random) $\pm 0.06$ (systematic) or 
$-0.72^{+0.22}_{-0.25}$ (random) $\pm 0.06$ (systematic)). However, when
we relax our 2-sigma rejection criteria to an iteratively obtained 3-sigma 
deviation from the fit, we find a metallicity dependence of 
$-0.71 \pm 0.17$ mag dex$^{-1}$, which agrees well with Shappee \& Stanek's 
(2011) values. We investigate the effects of the rejection criteria in
Figure 16, where we plot $\gamma_{VI}$ for various sigma rejection limits.
Here we find that the slope of the P-L intercept vs. metallicity
becomes systematically less steep with stricter rejection limits, and
is minimized at 2-sigma. We attribute this to the fact that stricter
rejection limits are more robust to outliers. Rejections stricter than 
2-sigma, however, produce results
that do not follow this trend, and are likely affected by random noise
due to low number statistics. Our sample size thus limits the effectiveness
of the sigma-clipping algorithm below 2-sigma, and as such, we use the 
2-sigma rejection limit for the final quoted result 
for this paper. The steeper metallicity dependence we find with
relaxed rejection constraints is an interesting result that may possibly
be due to blending in the more crowded inner regions of the galaxy. Tighter
constraints will help to remove blended stars from the
fit, thus producing a more reliable result. We further investigate the possible
effects of blending below.

Majaess et al. (2011) applied the Shappee \& Stanek's (2011) metallicity
correction of -0.80 mag dex$^{-1}$ to the Magellanic Clouds, and found that 
the distance to the galaxies obtained through this correction was in strong 
disagreement with those published through other methods, with no correction
producing much more consistent results. They therefore argue that the
VI Wesenheit P-L relation is insensitive to metallicity, and that Cepheid
metallicities therefore offer a negligible source of uncertainty to
derived extragalactic distances. They suggest that the steep
dependences on metallicity found in other studies are due to blending 
effects in the more crowded inner region of the galaxies, which would 
mimic a brightening of the P-L relation due to increased metallicity. 
This is consistent with the results of other authors such as Udalski et al. 
(2001), who also find no metallicity dependence of the P-L
relation when comparing its slope and intercept between three individual 
galaxies in the OGLE-II microlensing survey that are negligibly affected 
by blending. Mochejska et al. (2004) show that blending and crowding can 
have a significant effect by comparing the fluxes obtained for Cepheids in
ground-based and HST images, and that the problem is worse for more
distant galaxies where linear resolution is poorer.

If blending were an issue in our data, we would expect
to see an overall decrease in the average amplitude of the light curves
of Cepheids in the increasingly more crowded regions, as the blending
damps the observed light variations. We test this possibility by plotting 
the V-band amplitude of the Cepheids' light curves as a function of period in 
Figure 17, with the stars separated into their $\sim 1$ arcmin radial distance 
bins. As was also observed in the distance-separated P-L relations (Figures 9 
through 12), there is a larger proportion of short-period Cepheids at larger 
distances from the center, which could at least in part be due to the 
galaxy's radial metallicity gradient (Becker et al. 1977). The higher surface
brightness in the inner-most regions may also contribute to this effect,
as more of the fainter short-period Cepheids in the inner bins would have been
rejected due to low signal-to-noise than in the outer bins. 
We do observe an increase in amplitude with increasing radial 
distance only between the inner 1-2' and 2-3' bins, but not in the 
outer-most bins, where the upper-envelope of amplitudes is instead decreasing 
with increasing radial distance. This may be due to metallicity effects: 
some studies have shown that the amplitudes of longer-period Cepheids are 
larger in higher metallicity environments (van Genderen 1978, Paczynski \& 
Pindor 2000), which would explain the decreasing amplitude
with increasing radial distance in the outer distance bins in Figure 17. 
This implies that blending may be a factor only in the central one to two
radial bins. However, Szabados \& Klagyivik (2012) note that 
Cepheid amplitudes are smaller at higher metallicity, especially
for short-period Cepheids. Additionally, there are few high-amplitude 
short-period Cepheids in each of our distance bins, so the lack of these
types of stars in the outer-most bin may be due to statistical 
fluctuations. As such, the trends seen in Figure 17 may be
inconclusive. 

We can investigate this
further by observing the scatter in the P-L relations for each of the 
individual bins (Figures 9 through 12). If the amount of blending is
increasing toward the central, more crowded, regions of the galaxy, then
we would expect the scatter of the P-L relation to be progressively 
larger at smaller radial distances from the center. 
This is an effect that is readily apparent, for instance, in plots of the 
distance modulus derived from Cepheids at different galaxy radii for
the M33 and M106 data in Majaess, et al. (2009). There the scatter is
significantly larger for Cepheids in the inner region of the galaxy than
those in the outer region. In our M101 images, the one-sigma scatter 
of the data from the fit for each of the distance bins (from the inner-most to
the outer-most) is 0.100, 0.157, 0.107, and 0.093 mag.
Although the outer-most distance bin has the lowest scatter,
there is otherwise no apparent trend with radial distance.
However, if we relax the rejection constraints from 2-sigma to 3-sigma,
we see a one-sigma scatter of the data from the fit (from the inner-most to
the outer-most distance bin) of 0.202, 0.203, 0.188, and 0.125 mag. 
As with the Majaess, et al. (2009) data, this demonstrates a progressively 
tighter P-L relationship outside 3 arcminutes radius.
If the additional scatter
in the inner regions is due to blending, it could create a systematic
error in magnitude that is too bright with decreasing distance, and thus
result in a metallicity dependence that is too steep. This is a possible
explanation for the steeper metallicity dependence that we calculate from
the more relaxed rejection levels. The relative lack of this trend in the
2-sigma rejected data further suggests that the stricter rejection
criterion is more effective at removing blending effects, and is thus
more robust. This suggests that blending has a negligible effect on
our 2-sigma result, and we proceed with that result accordingly. 

The difference in amplitude and scatter of our Cepheids with 
different rejection constraints suggests that blending
is an issue, and some correction must be applied to Cepheid studies to
account for this. We reject outliers through a sigma-clipping algorithm
here to minimize the effects of blending, but other corrections can
be applied to each star based on the assumed effects of crowding, such as 
that developed by Riess et al. (2009). Whatever method is used, it is 
clear that extra caution should be exercised in the inner crowded fields
of extragalactic Cepheid studies.

The dependence of the P-L zero-point on metallicity derived in any study is
also highly sensitive to the adopted radial metallicity gradient. 
There has been significant contention, for instance, on the value of the
metallicity gradient in NGC 4258 (Bresolin 2011, Bono et al. 2008), which
significantly impacts conclusions on the metallicity dependence of the
P-L relation. This is therefore another possible source of error which
could be minimized with the acquisition of more accurate metallicity 
measurements.

\section{Improving Cepheid Distances by Observing in the Infrared}

Freedman et al. (2011) show that 
the P-L relation from 3.6 $\mu m$ Spitzer observations
have minimal systematic effects due to both extinction and metallicity.
They calculate that improvements in the extragalactic distance scale
calibration by using infrared Cepheid relations can allow future work
to improve the 10\% uncertainty on the Hubble Constant found in the 
Key Project (Freedman, et al. 2001) to less than 2\%. As with the
optical, however, crowding can also be an issue in the infrared. 

To address known systematics in an attempt to derive an increasingly
precise and accurate value of the Hubble constant the Carnegie Hubble
Program (Freedman et al. 2012) has been using Spitzer to observe known
Cepheids in the mid-infrared. This has led to two new realizations, one
concerning the dominant source of crowding in the mid-IR and the other
concerning the sensitivity of select mid-IR bands to metallicity.

The (confirmed) expectation about crowding in the near and mid-infrared
was that the OB-star populations, that are coeval and generally co-located 
with Cepheids, are less important sources of crowding as one moves to 
longer wavelengths where the intrinsically redder Cepheids become 
brighter than the bluer OB stars. However, a second population of 
contaminating sources come to the fore-front, especially at mid-IR 
wavelengths: the dust-enshrouded and intrinsically red (but still very 
luminous) asymptotic giant branch (AGB) stars. More uniformly spread over 
the field, the intermediate-aged AGB star population quickly becomes the 
dominant (new) source of contamination for Cepheids. At the largest 
distance at which Cepheids have been currently discovered the AGB population 
severely limits the use of Spitzer; only much higher resolution 
observations will overcome this confusion-limiting factor.

As shown in Monson et al. (2012) and Scowcroft et al. (2011) the use of
mid-infrared bands (specifically the IRAC 3.6 and 4.5$\mu$m of Spitzer) 
can be used to further reduce the effects of reddening in determining 
true distance moduli to Cepheids in nearby galaxies. One surprising
result was that the 4.5$\mu$m band is singularly sensitive to metallicity,
for the longest-period (coolest) Cepheids. This is due to the presence of
a CO band head lying across the 4.5$\mu$m region of the spectrum. 
The 3.6$\mu$m filter is unaffected by CO and appears to be
measuring pure continuum radiation. Thus the anticipated advantages of
moving as far into the mid-infrared as possible (to reduce the systematic 
effects of reddening) are met at 3.6$\mu$m, but complicated by metallicity 
effects in the 4.5$\mu$m band. However, it should be noted that the CO 
is effectively gone (the molecule is dissociated) from atmospheres of 
shorter-period (hotter) Cepheids, should future studies need to use this 
band and these hotter Cepheids for distance determinations.

\acknowledgments

This research has made use of the NASA/IPAC Extragalactic Database
(NED) which is operated by the Jet Propulsion Laboratory, California
Institute of Technology, under contract with the National
Aeronautics and Space Administration. This research has also made use of
NASA's Astrophysics Data System. We greatly appreciate the contributions
of the referee, who made many suggestions that improved the overall
quality of this paper.

{\it Facilities:} \facility{HST (ACS)}

\clearpage
\begin{figure}
\centerline{\includegraphics[width=15cm]{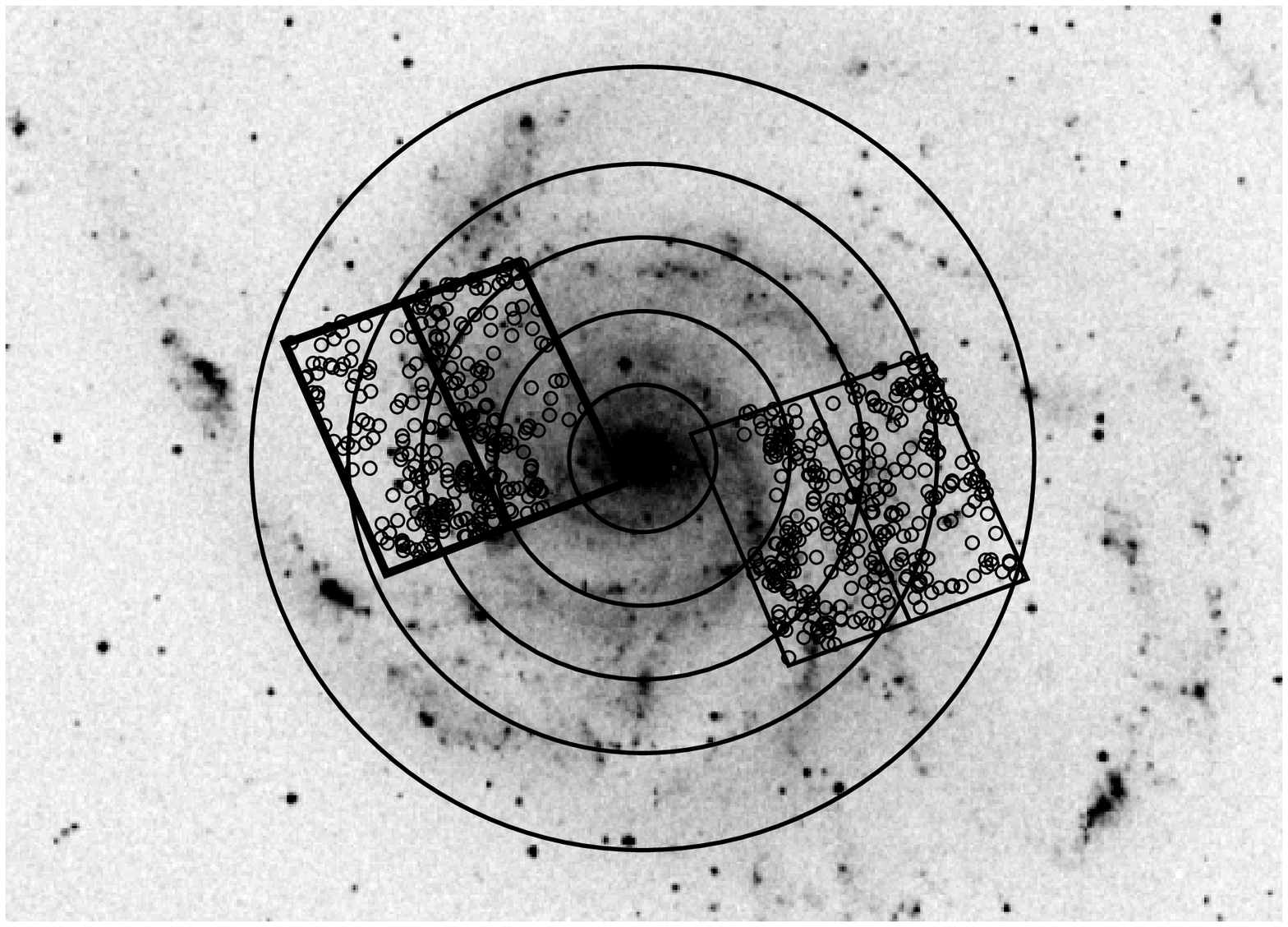}}
\figcaption[f1.ps]{A DSS image of M101, overlaid with the locations
of the ACS images of Field 1 (thick-lined) and Field 2 
(thin-lined). Small
circles mark the location of the Cepheid candidates found in this analysis.
The large circles centered on the galaxy represent the annuli used 
to bin the data in order to study the effects of metallicity on the 
Cepheid period-luminosity relationship.}
\end{figure}

\clearpage
\begin{figure}
\centerline{\includegraphics[width=15cm]{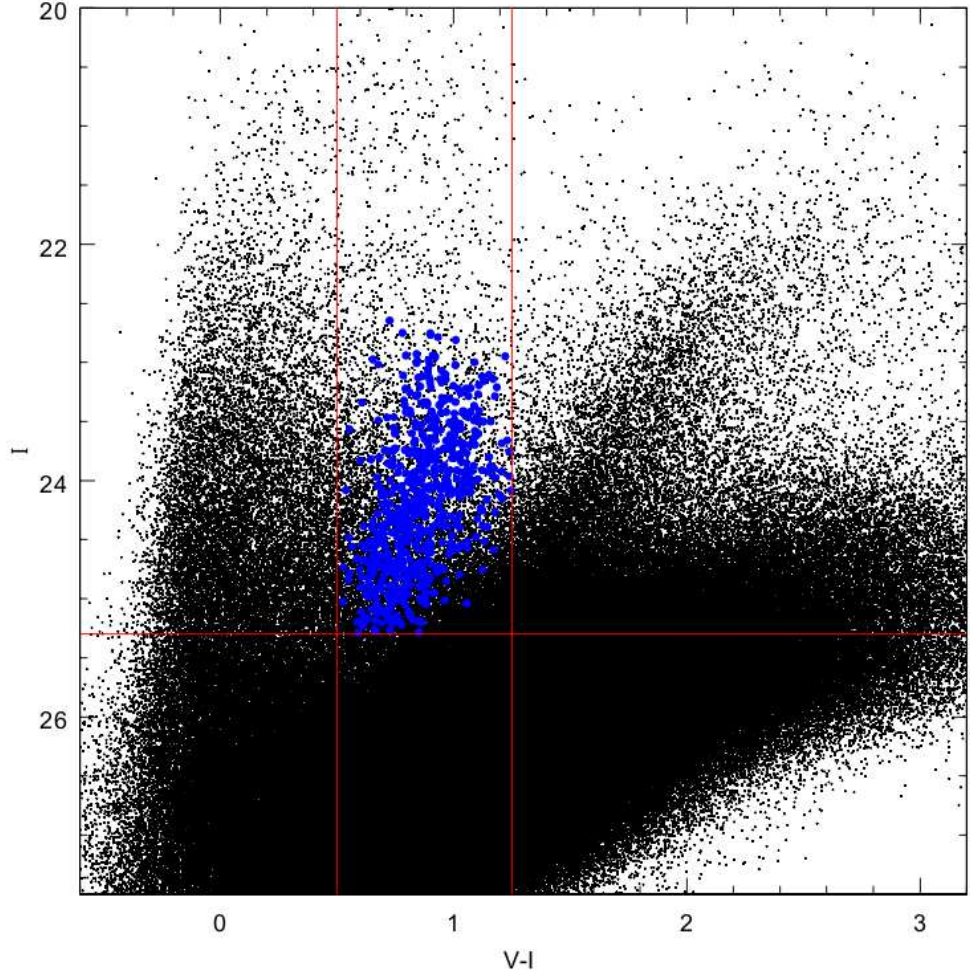}}
\figcaption[f2.eps]{I-band magnitude vs. V-I color for all stars
identified in both fields of M101 (black points). Blue circles mark the 
Cepheid candidates. Red lines show the magnitude and color cutoffs used in
the Cepheid selection.
}
\end{figure}

\clearpage
\begin{figure}
\centerline{\includegraphics[width=15cm]{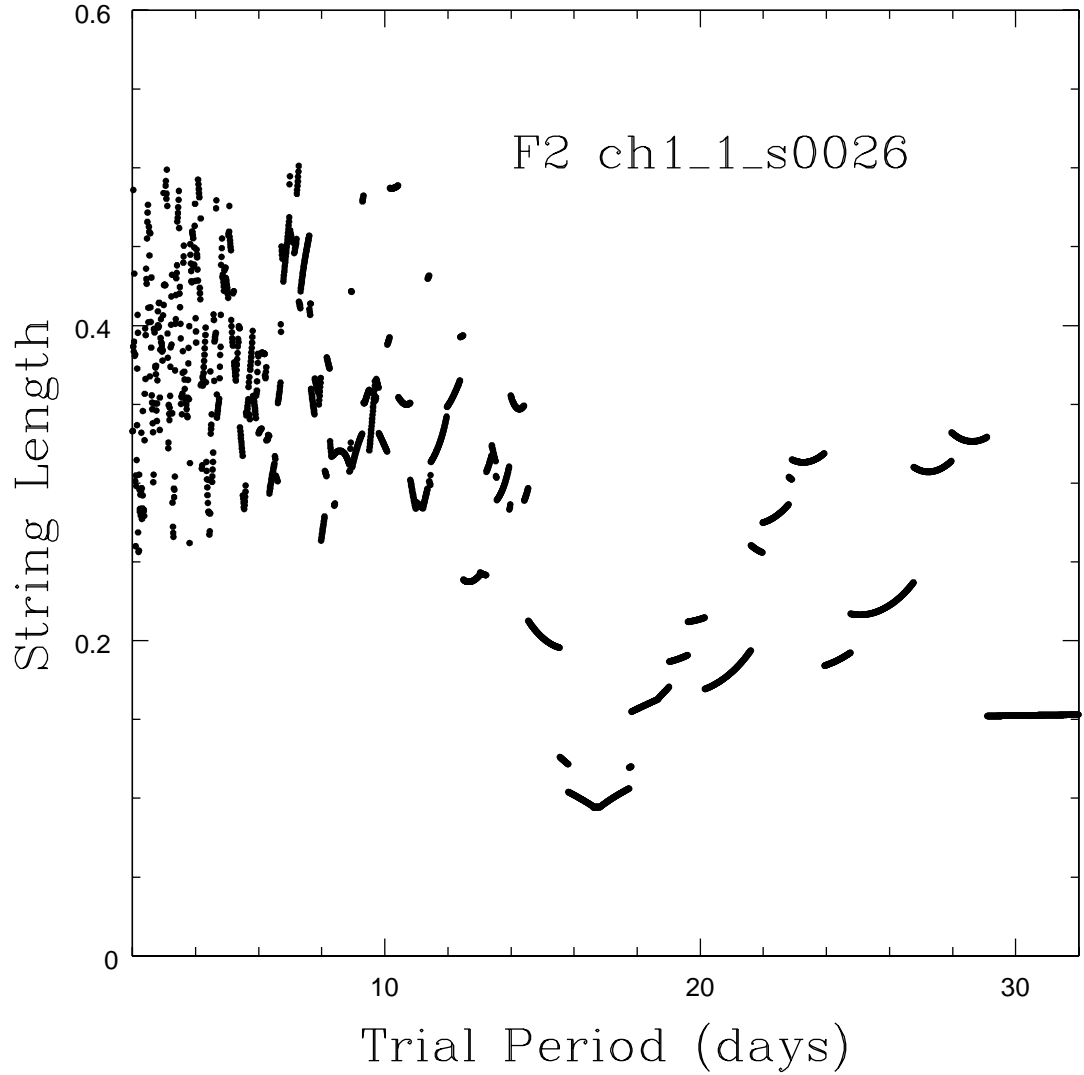}}
\figcaption[f3.ps]{Example string length vs. trial period for one star.
The minimum string length value corresponds to the true period (16.60 days
in this case). The light curve and phase diagram for this star is presented
in Figure 4.
}
\end{figure}

\clearpage
\begin{figure}
\centerline{\includegraphics[width=15cm]{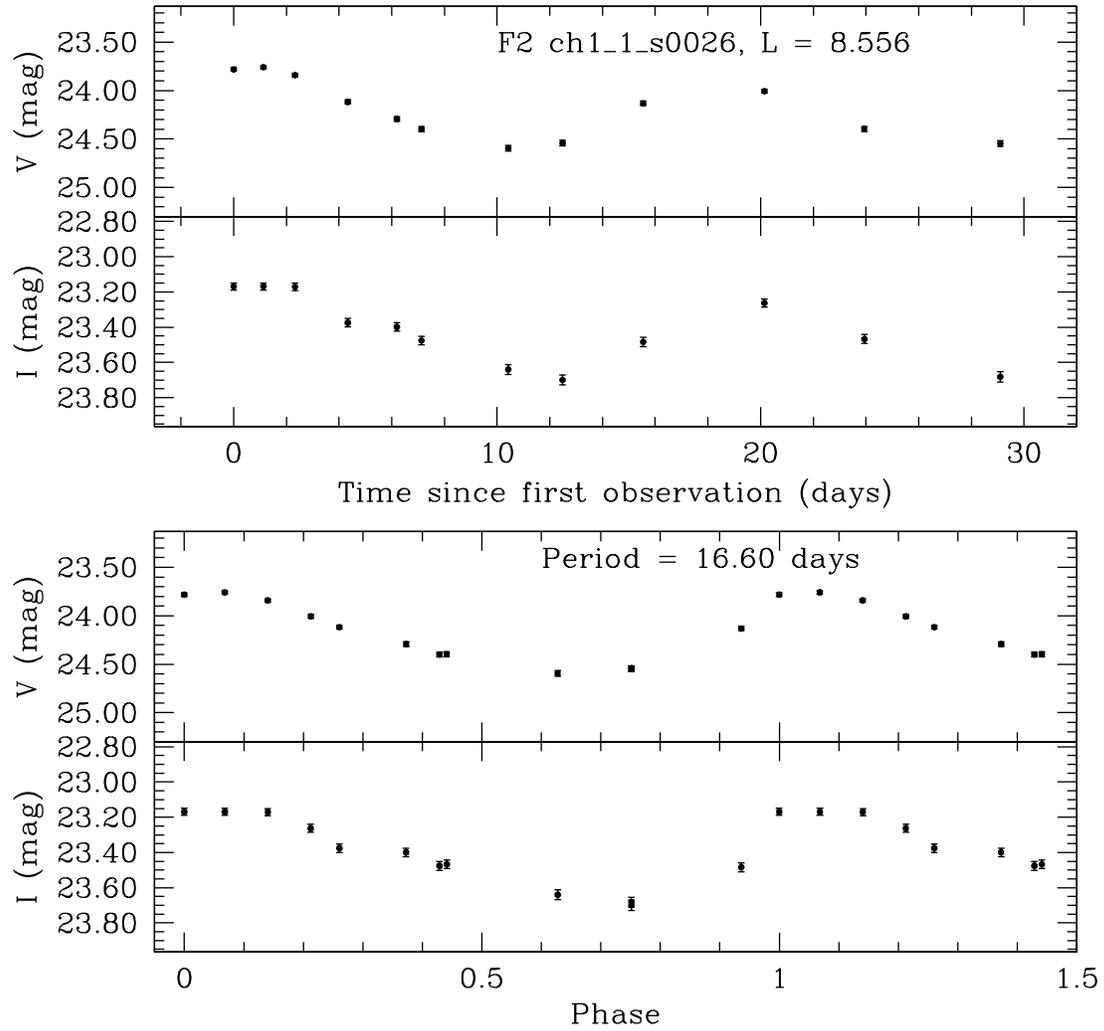}}
\figcaption[f4.ps]{{\bf Top:} Observed time sequence within the 30-day 
observing window for one star for both the V and I bands. 
{\bf Bottom:} Phase-folded light curve for the same star using the period 
obtained with the string length minimization technique, as shown in Figure 3. 
}
\end{figure}

\clearpage
\begin{figure}
\centerline{\includegraphics[width=15cm]{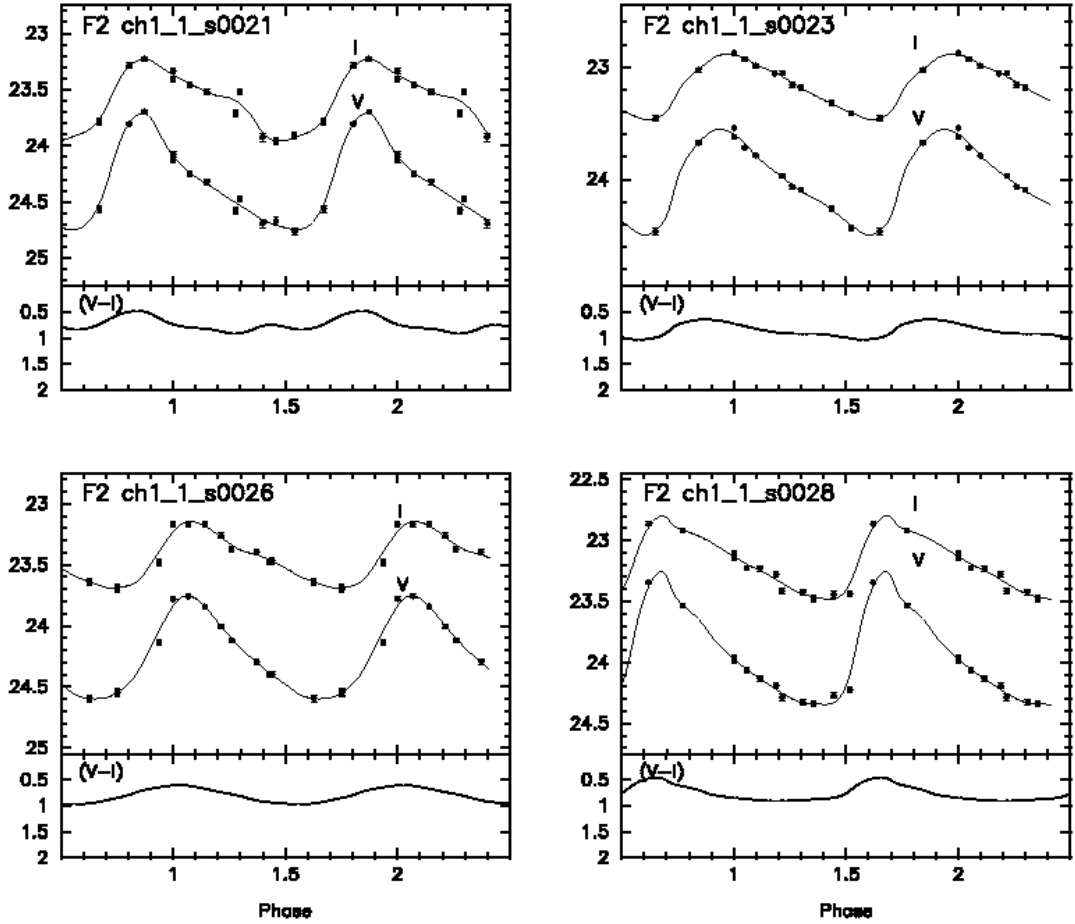}}
\figcaption[f5.ps]{Phase-folded light curves for four stars with fits
overlaid on the data points. The lower left panel shows the fit for
the star used in Figures 3 and 4. Light curves for all stars are available in
the online edition.
}
\end{figure}

\clearpage
\begin{figure}
\centerline{\includegraphics[width=15cm]{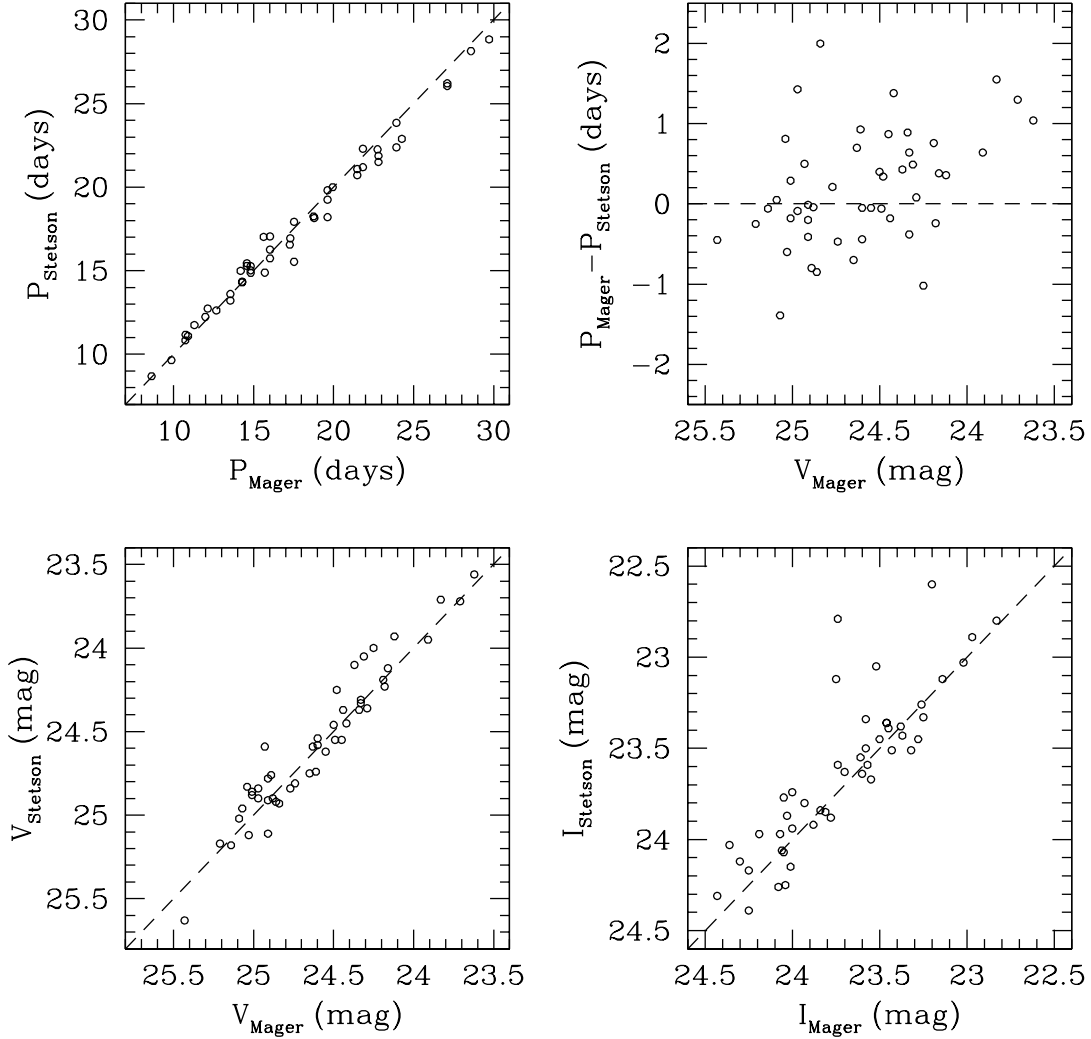}}
\figcaption[f6.ps]{Comparison of the Cepheid periods ({\bf top left}), and 
magnitudes ({\bf bottom left}: V, {\bf bottom right}: I) measured in this 
project to those of Stetson et al. (1998). {\bf Top right}: Residuals of 
the periods between the two samples as a function of our V-band magnitudes. 
There is no systematic difference, with a median residual of 0.07 days, and a 
standard deviation of 0.74 days.
}
\end{figure}

\clearpage
\begin{figure}
\centerline{\includegraphics[width=15cm]{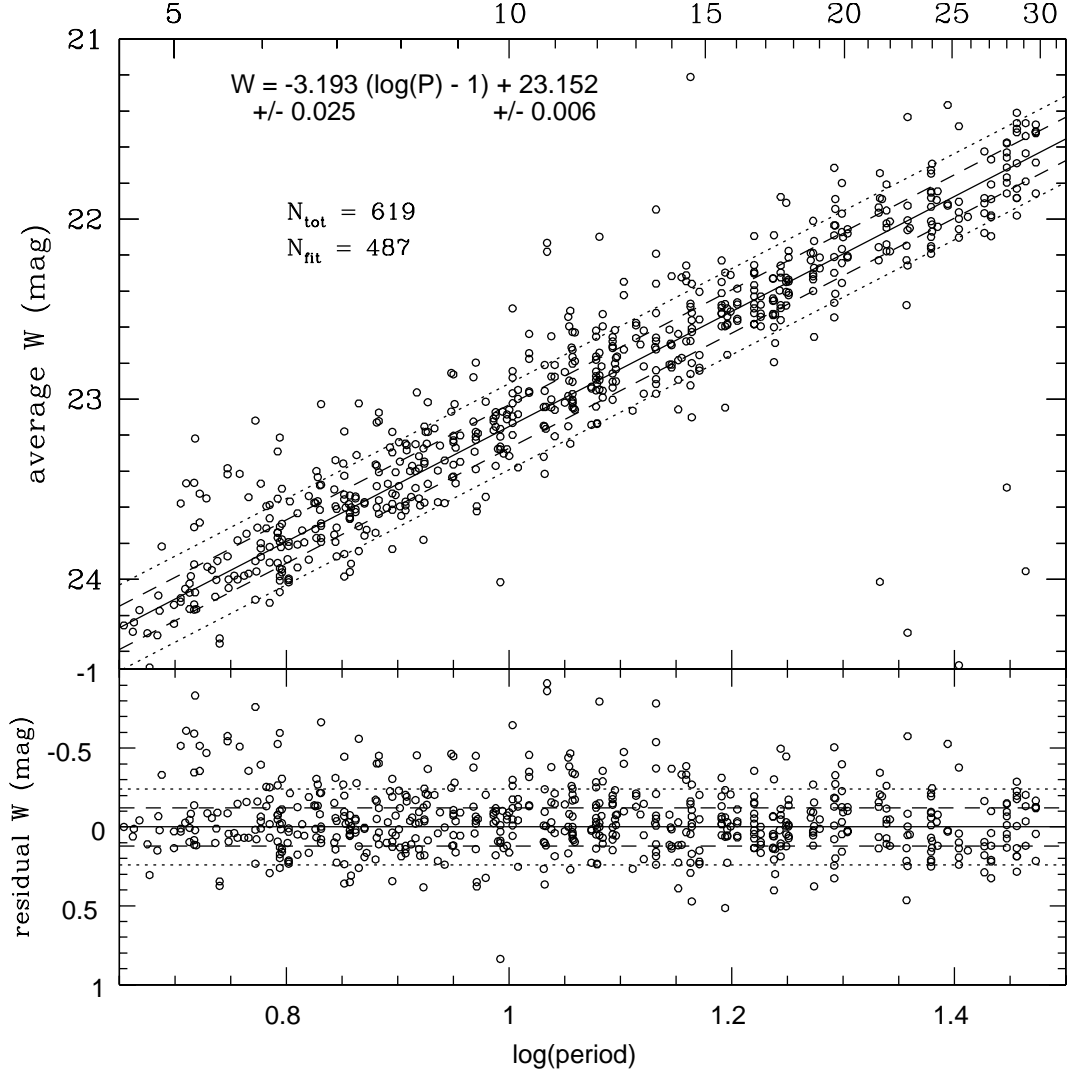}}
\figcaption[f7.ps]{{\bf Top panel:} Reddening-free average Wesenheit 
magnitude of all Cepheid
candidates in M101 vs. the log of their periods. The solid line shows
the linear least-squares fit to the data, the dashed lines show the 
1-sigma standard deviation of the data to the fit, and the dotted lines 
show the 2-sigma deviation. N$_{tot}$
is the total number of Cepheids, and N$_{fit}$ is the
number of Cepheids used to calculate the fit, after the iteratively 
determined 2-sigma rejection limit was applied. {\bf Bottom panel:} 
Residual Wesenheit magnitude of each Cepheid from the fit to the data,
with 1-sigma and 2-sigma standard deviations shown as described for the 
top panel.
}
\end{figure}

\clearpage
\begin{figure}
\centerline{\includegraphics[width=15cm]{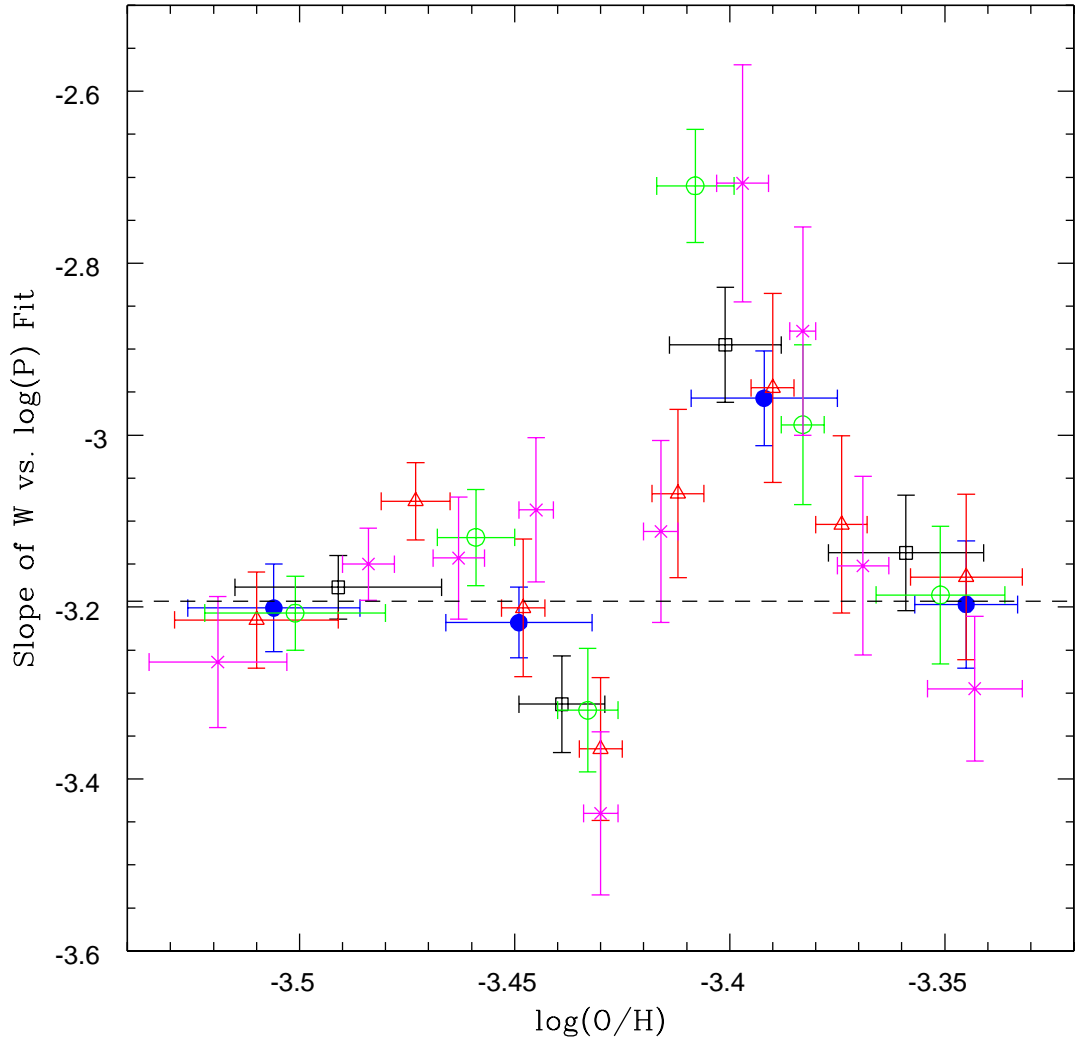}}
\figcaption[f8.ps]{Dependence of the slope of the fit to the Cepheid
P-L relation on metallicity for several samples chosen in different ways. 
Error bars show the standard deviations of
the metallicity of the Cepheids in each annulus, and the uncertainties
on the calculated slopes. There is no clear dependence, and so we fix the
slope (dotted line) to that measured from the entire data-set (Eq. 10).
}
\end{figure}

\clearpage
\begin{figure}
\centerline{\includegraphics[width=15cm]{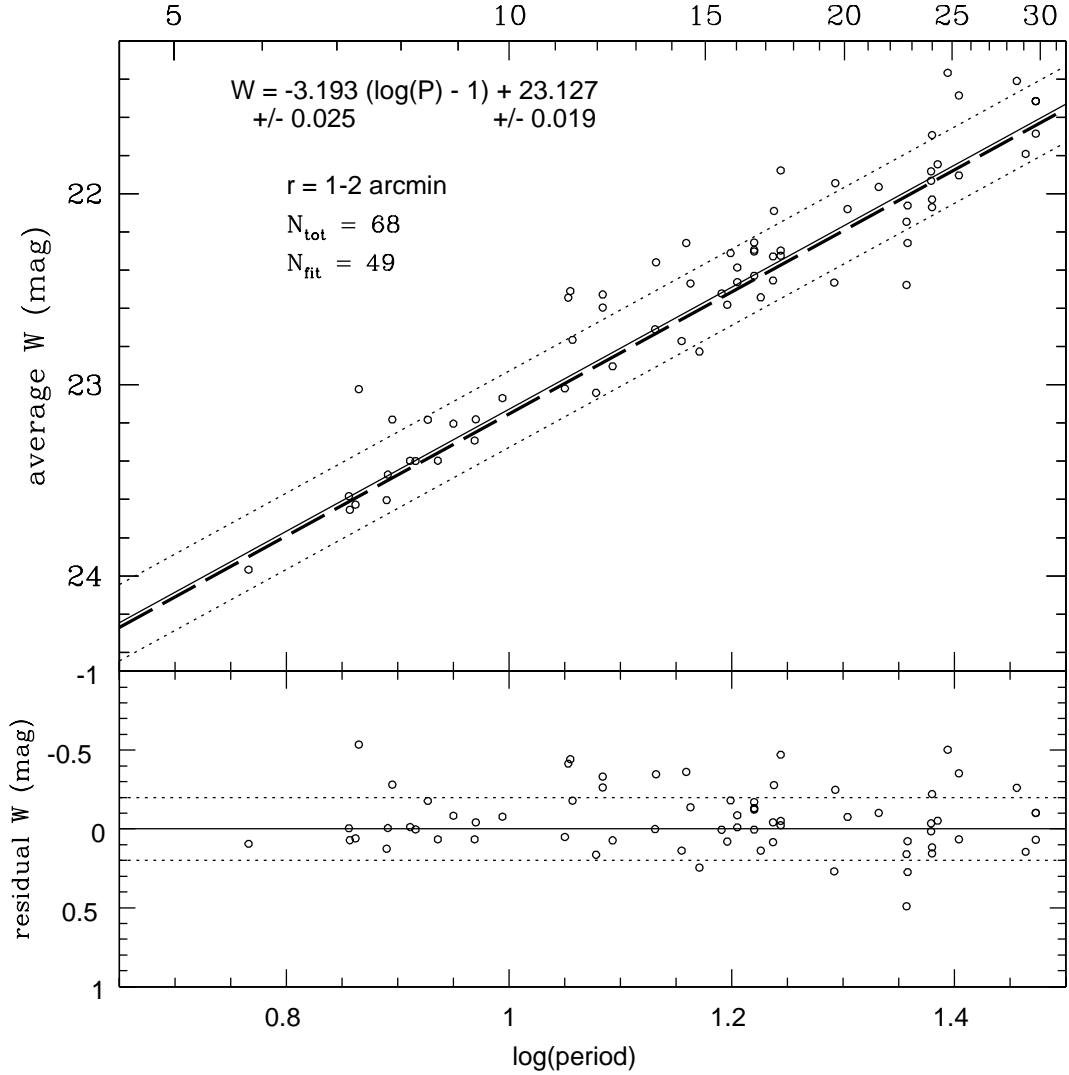}}
\figcaption[f9.ps]{{\bf Top panel:} Reddening-free average Wesenheit magnitude of the Cepheids
in the first distance bin ($1-2\arcmin$ from the center of the galaxy) 
vs. the log of their periods. A linear least-squares fit was 
performed with the slope fixed to that calculated from the entire dataset. 
The solid line shows the fit to the data, and the dotted lines show the
2-sigma deviation of the data to the fit. N$_{tot}$
is the total number of Cepheids in the distance bin, and N$_{fit}$ is the
number of Cepheids used to calculate the fit, after the iteratively 
determined 2-sigma rejection limit was applied.
The thick long-dashed line is the fit to the entire data-set, as in Figure 7.
{\bf Bottom panel:} Residual Wesenheit magnitude of each Cepheid from the fit 
to the data, with 2-sigma standard deviations shown as described for the
top panel. 
}
\end{figure}

\clearpage
\begin{figure}
\centerline{\includegraphics[width=15cm]{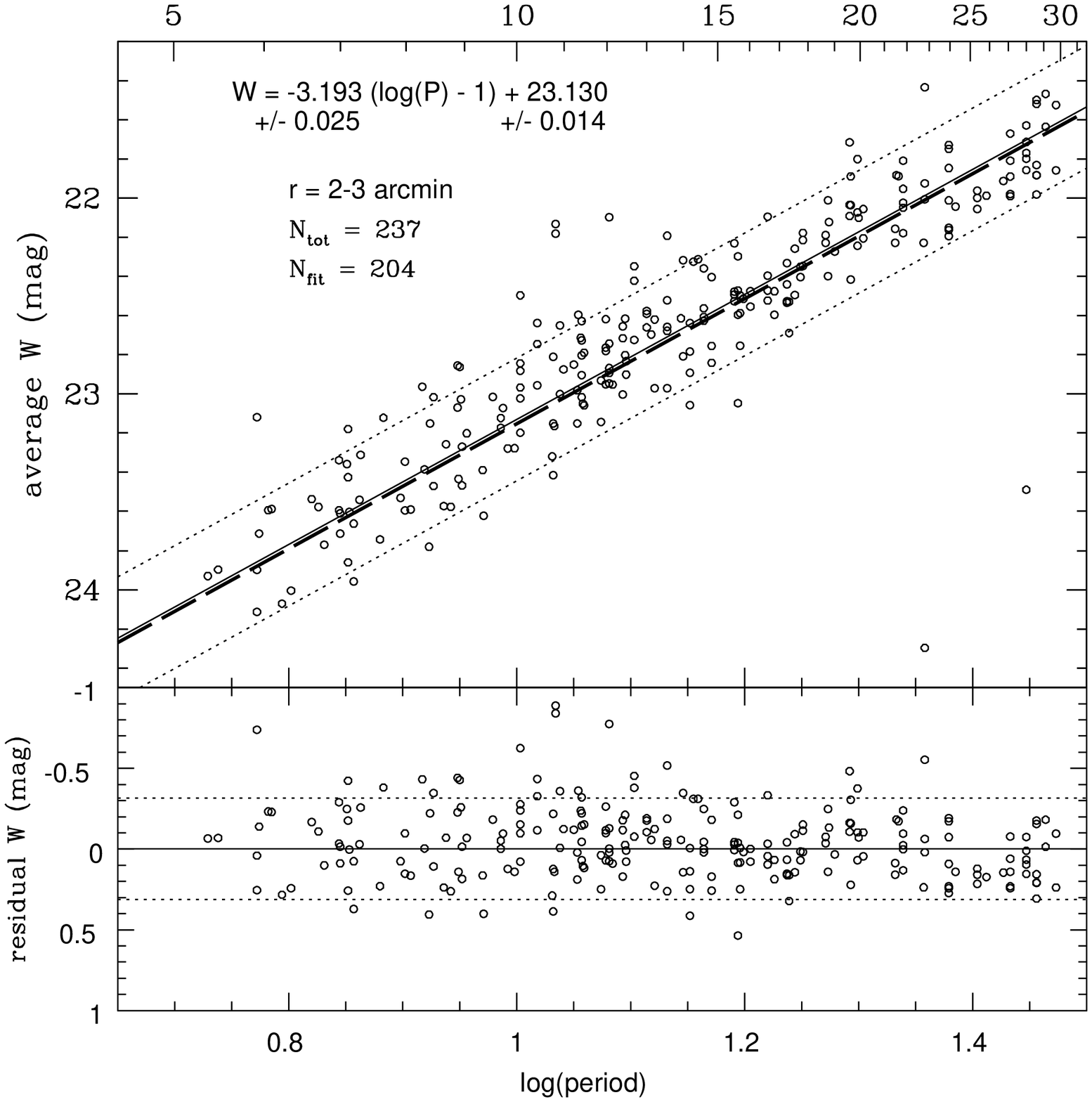}}
\figcaption[f10.ps]{Period-Luminosity relation for the second distance
bin; the same as Figure 9, only for Cepheids within $2-3\arcmin$ of 
the center of the galaxy.
}
\end{figure}

\clearpage
\begin{figure}
\centerline{\includegraphics[width=15cm]{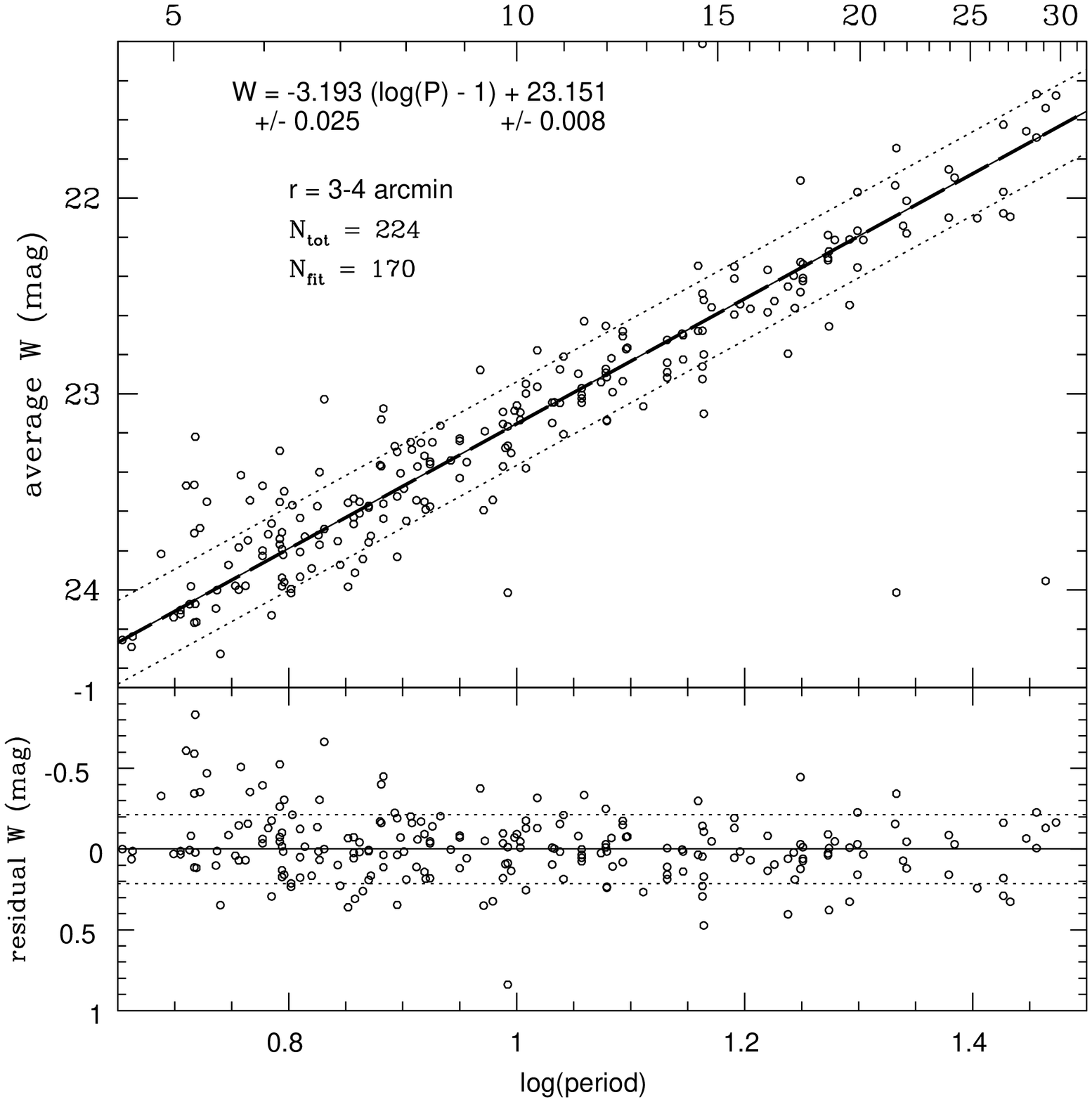}}
\figcaption[f11.ps]{Period-Luminosity relation for the third distance
bin; the same as Figure 9, only for Cepheids within $3-4\arcmin$ of 
the center of the galaxy.
}
\end{figure}

\clearpage
\begin{figure}
\centerline{\includegraphics[width=15cm]{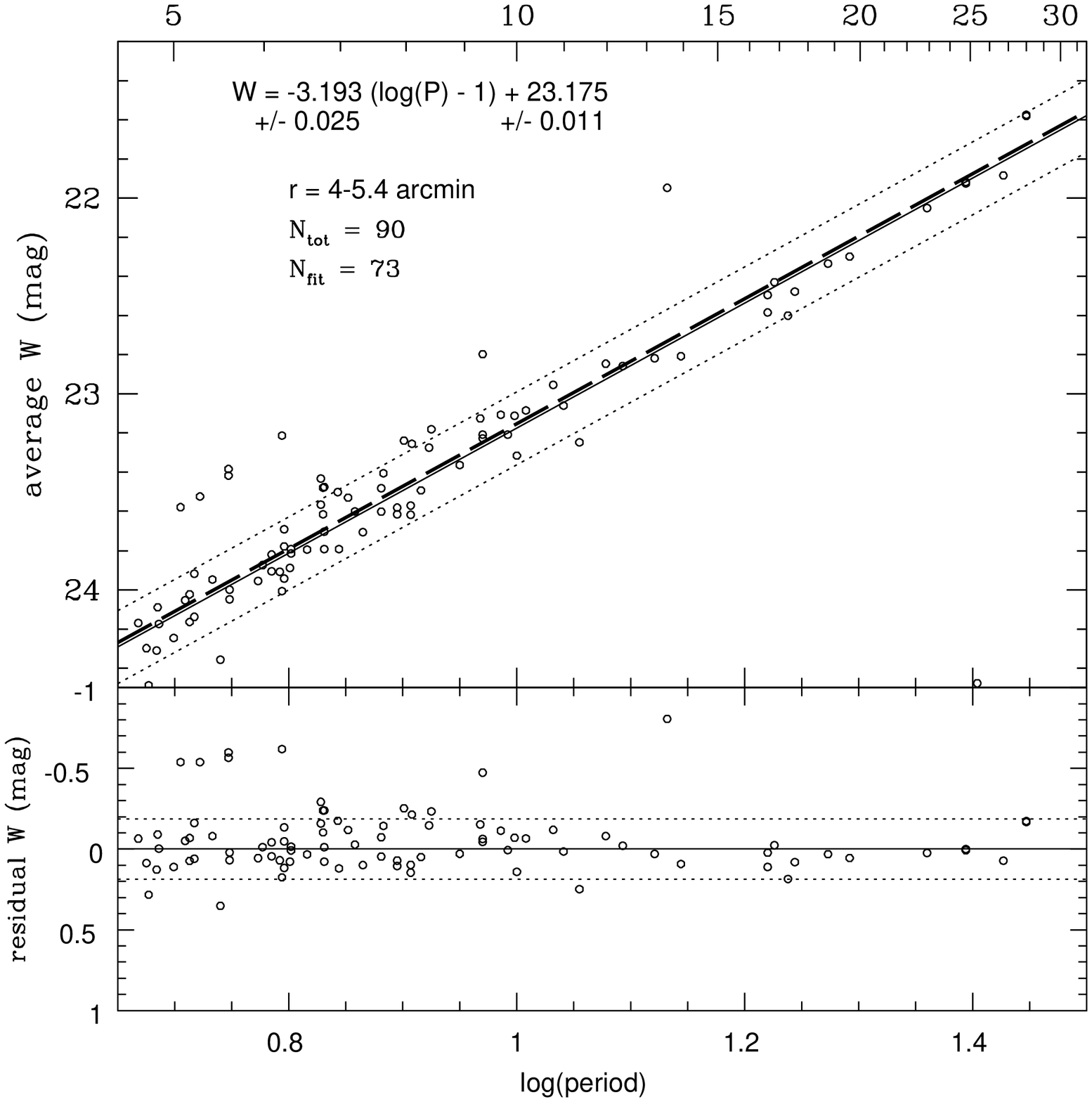}}
\figcaption[f12.ps]{Period-Luminosity relation for the fourth distance
bin; the same as Figure 9, only for Cepheids within $4-5.4\arcmin$ of 
the center of the galaxy.
}
\end{figure}

\clearpage
\begin{figure}
\centerline{\includegraphics[width=15cm]{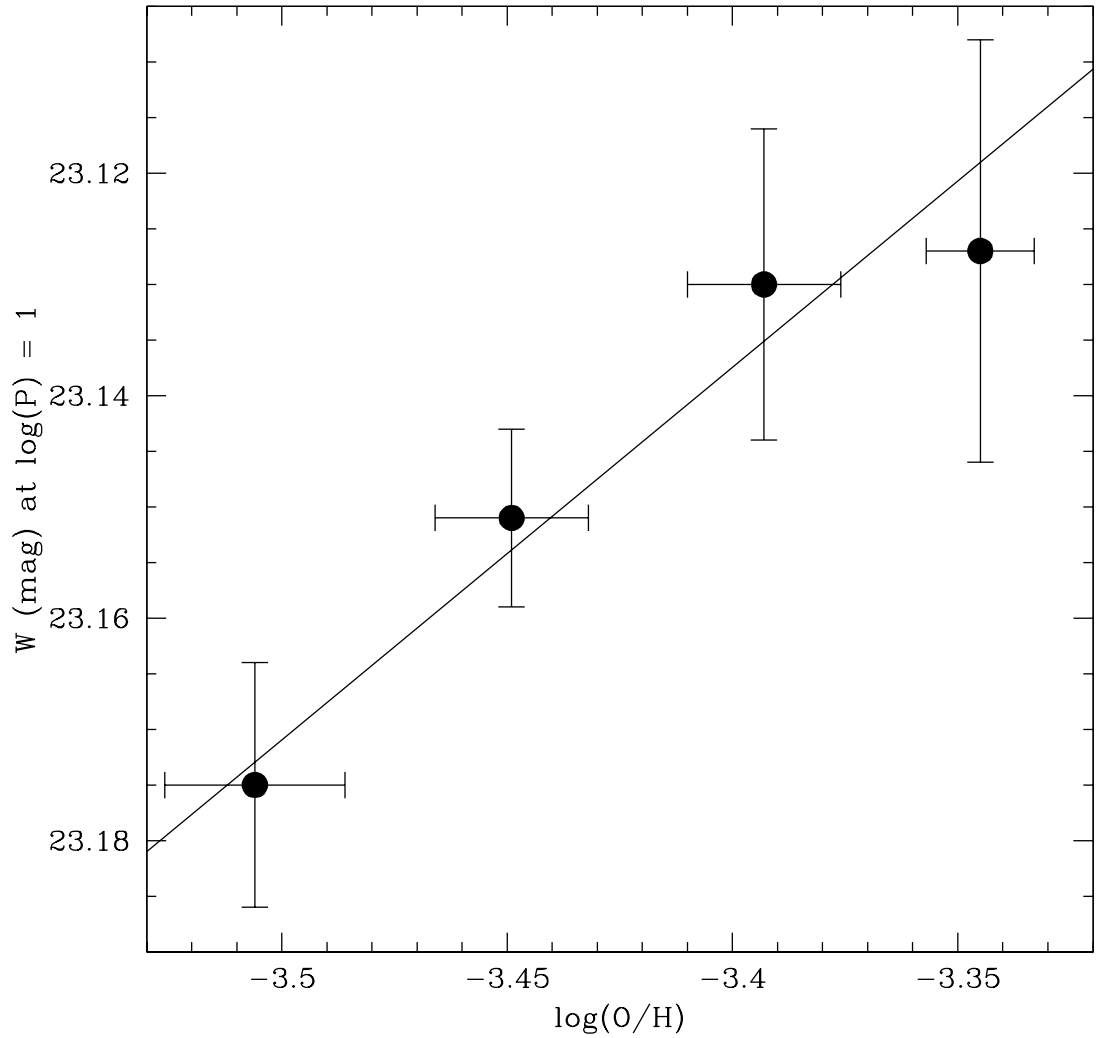}}
\figcaption[f13.ps]{Dependence of the intercept of the fit to the
Cepheid P-L relation at log(P) = 1 on metallicity. There is a 
significant trend of increasing brightness with increasing
metallicity (corresponding to decreasing distance from the galaxy
center). The weighted linear least-squares fit to the data is shown, with
a slope of $\gamma_{VI} = -0.33 \pm 0.12$ mag dex$^{-1}$.
}
\end{figure}

\clearpage
\begin{figure}
\centerline{\includegraphics[width=15cm]{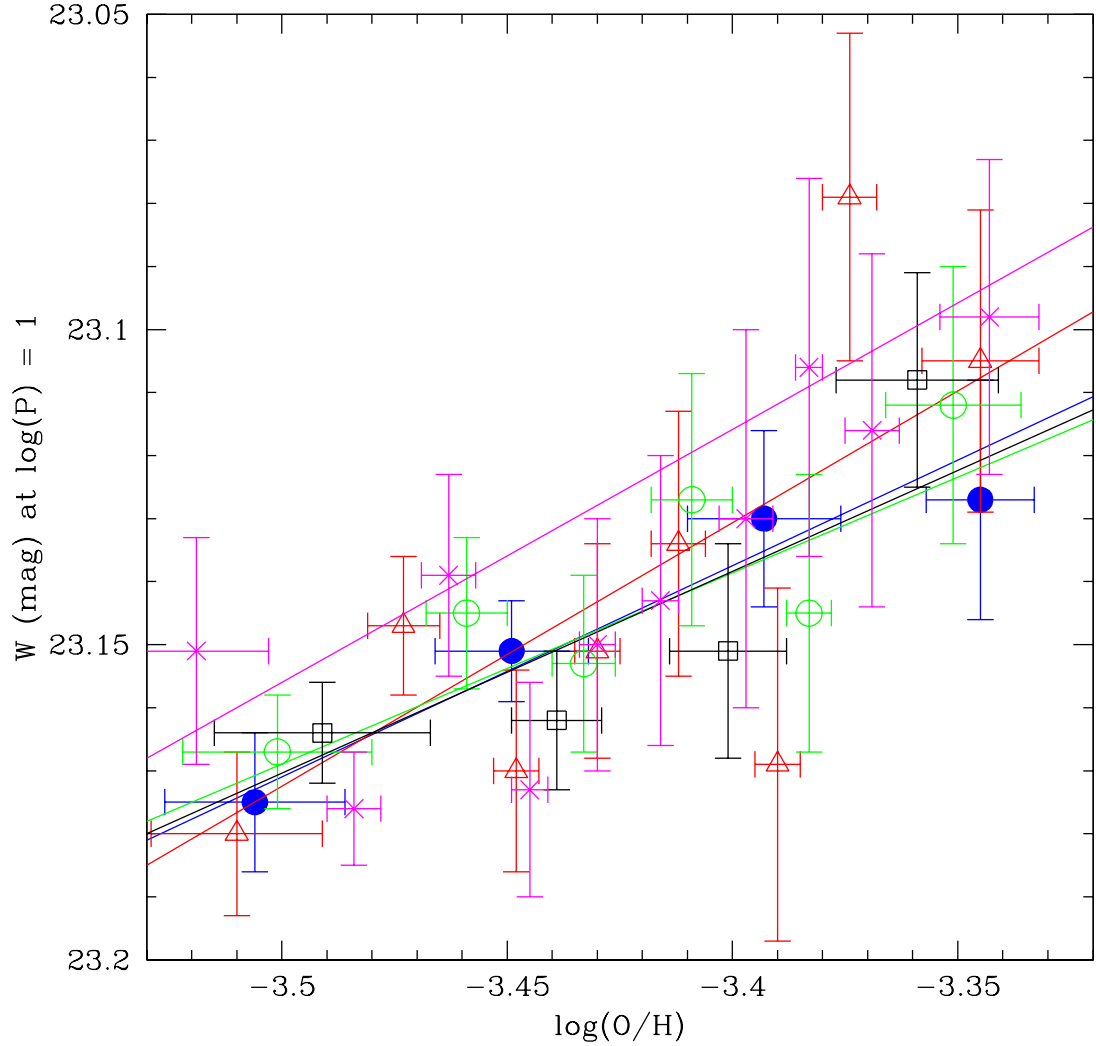}}
\figcaption[f14.ps]{Dependence of the intercept of the fit to the
Cepheid P-L relation at log(P) = 1 on metallicity for several
samples chosen in different ways. The weighted linear least-squares
fit to each dataset is shown with the same color as the corresponding
data-points. There is no significant dependence of the result on 
sample selection, with the slope of the weighted fits agreeing within
the errors. 
}
\end{figure}

\clearpage
\begin{figure}
\centerline{\includegraphics[width=15cm]{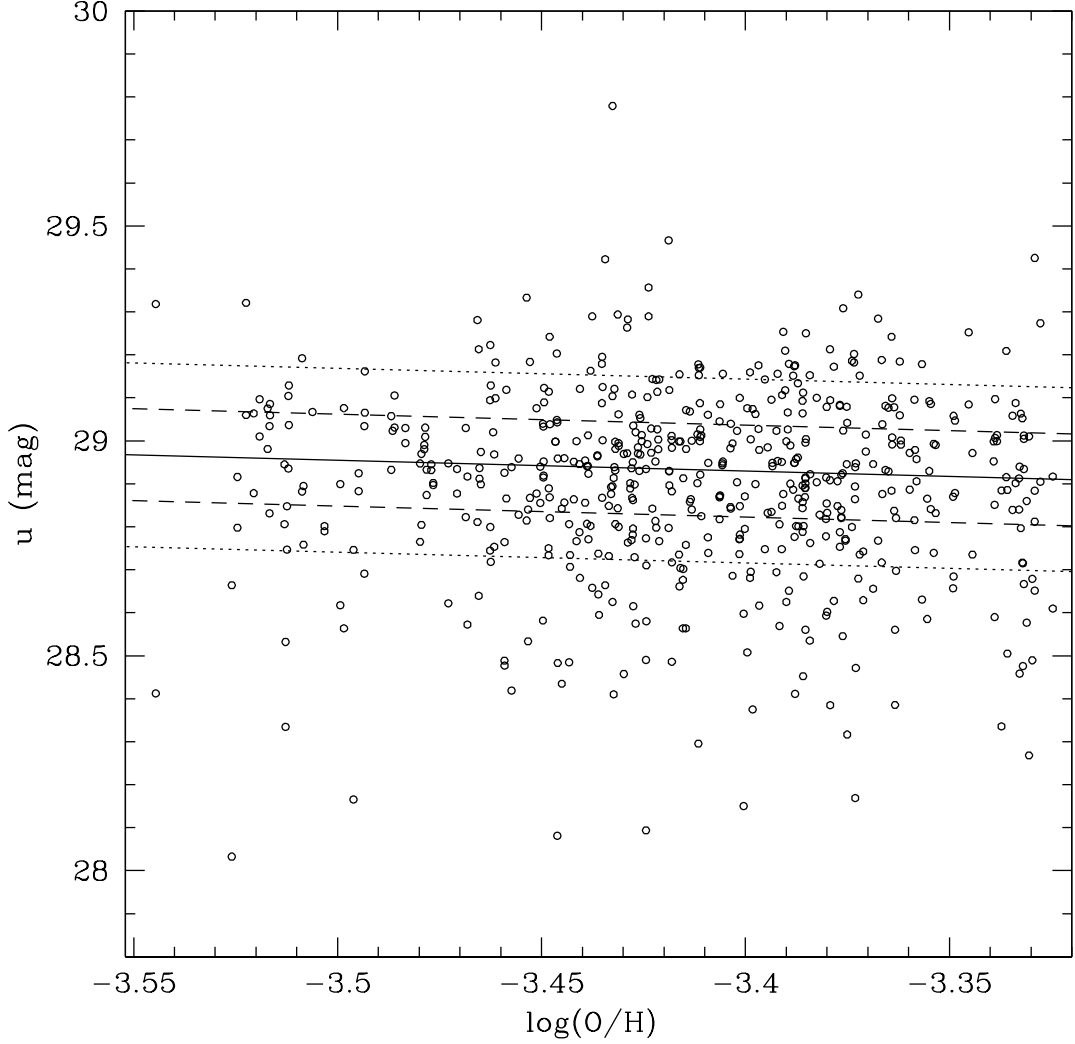}}
\figcaption[f15.ps]{The distance modulus for Cepheids in our sample 
as a function of metallicity, derived from the Wesenheit P-L relation
of the LMC with an assumed distance modulus of 18.48 mag
(Freedman et al. 2012). The solid line
is the iterative linear least-squares fit to the data, with the dashed line 
representing the 1-sigma standard deviation of the data from this fit, and
the dotted line representing the 2-sigma deviation. No data beyond the
iteratively determined 2-sigma
deviation were included in the fit. We correct for metallicity in our 
determination of the distance modulus to M101 by defining it to be the value 
of this fit at the metallicity of the LMC (log(O/H) = -3.5).  
}
\end{figure}

\clearpage
\begin{figure}
\centerline{\includegraphics[width=15cm]{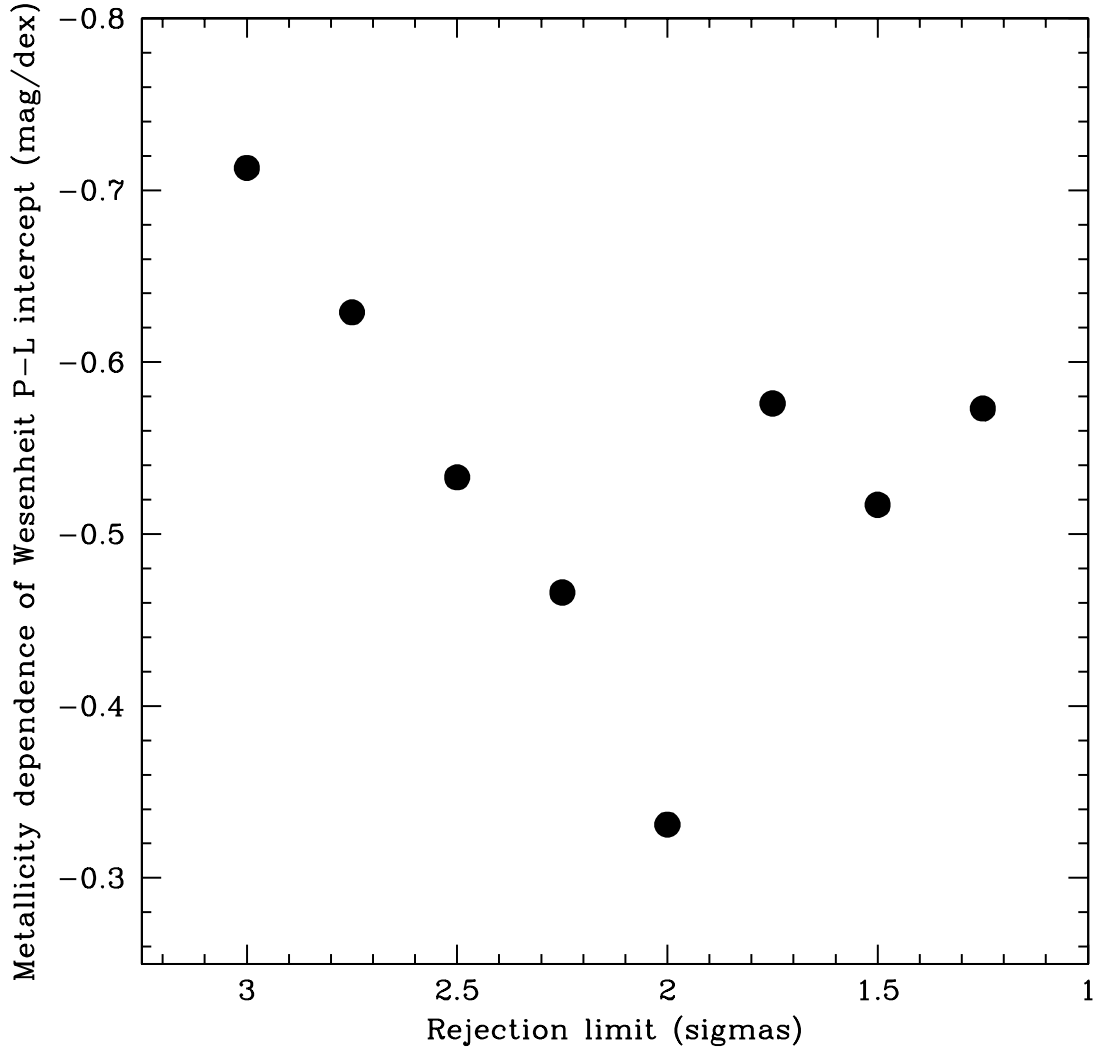}}
\figcaption[f16.ps]{The dependence of $\gamma_{VI}$ on the sigma
rejection limit applied to the data. Stricter rejection constraints
progressively produce a more mild dependence of the intercept of
the Wesenheit P-L relation on metallicity until a minimum value for
$\gamma_{VI}$ is obtained at the 2-sigma rejection level. The spurious results
for rejection limits less than 2-sigma are likely affected by low-number
statistics. 
}
\end{figure}

\clearpage
\begin{figure}
\centerline{\includegraphics[width=15cm]{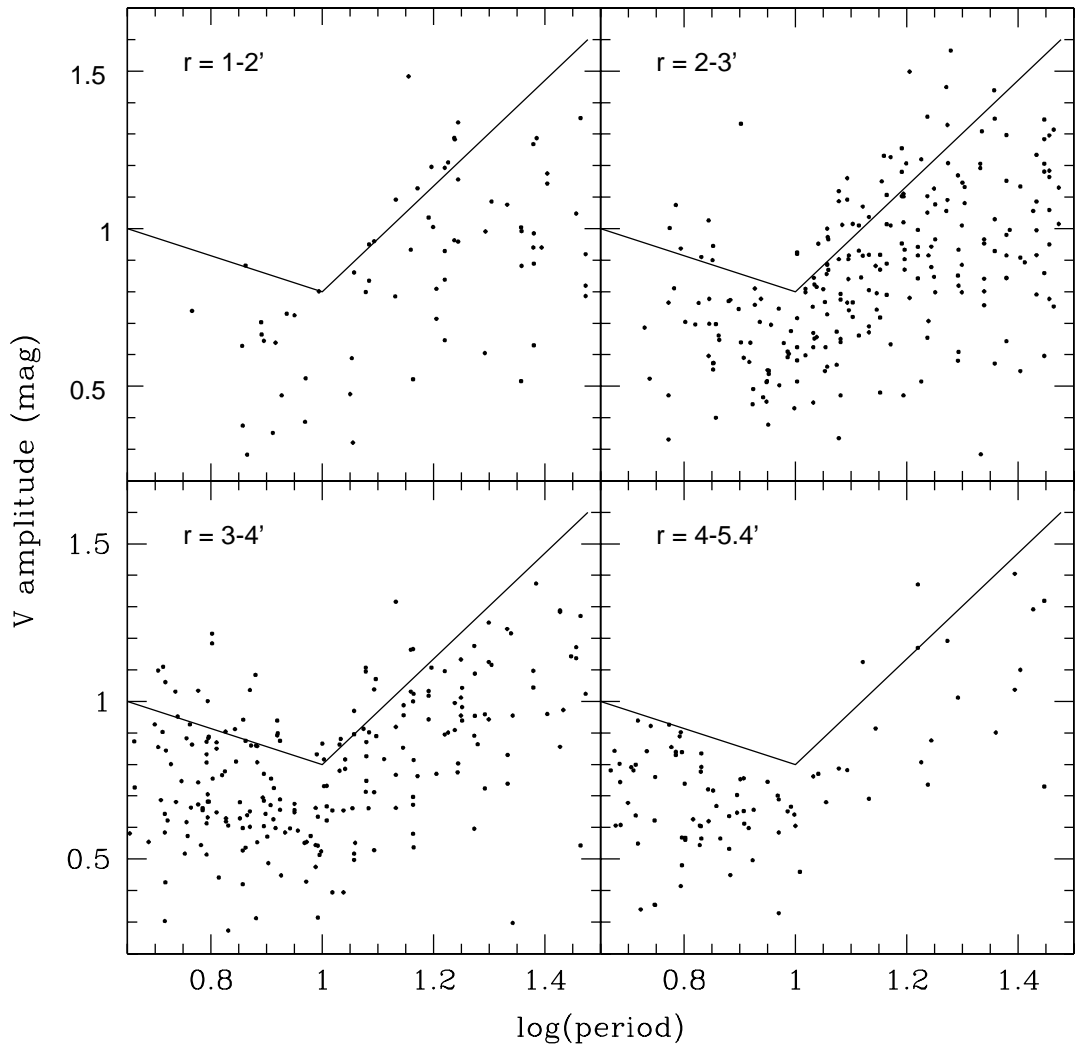}}
\figcaption[f17.ps]{The V-band amplitude of the Cepheids' light curves
as a function of period, with the stars in each distance bin separated
into different panels as indicated in the figure. The same solid lines 
are drawn in each of the panels to guide the eye in comparing the amplitudes
as a function of radial distance. As also observed in
Figures 9 through 12, there is a larger proportion of short-period 
Cepheids at larger radial distances, which is due at least in part to 
metallicity
effects. There is only a decrease in amplitude with decreasing radial 
distance between the inner two bins, which would be expected if the more 
crowded inner region is more strongly affected by blending.
}
\end{figure}

\clearpage
\begin{deluxetable}{llllllllllll}
\tabletypesize{\scriptsize}
\tablecolumns{12}
\tablecaption{Measured Quantities for Candidate Cepheids\label{Table-1}}
\tablewidth{0pt}
\tablehead{
\colhead{ID} & \colhead{RA} & \colhead{DEC} & 
\colhead{V} & \colhead{V$_{err}$} 
& \colhead{I} & \colhead{I$_{err}$} & 
\colhead{Period}
& \colhead{V$_{amp}$} & \colhead{I$_{amp}$} 
& \colhead{r} & \colhead{[O/H]}\\
& \colhead{(deg)} & \colhead{(deg)} & \colhead{(mag)} & \colhead{(mag)}
& \colhead{(mag)} & \colhead{(mag)} & \colhead{(days)} & \colhead{(mag)}
& \colhead{(mag)} & \colhead{(\arcmin)} 
}
\startdata
F1 ch1\_1\_s0001 & 210.92486 & 54.377676 & 24.341 &  0.037 & 23.522 &  0.022 & 18.75 & 1.192 & 0.676 &  4.605 & -3.519 \\
F1 ch1\_1\_s0002 & 210.91499 & 54.351110 & 24.250 &  0.039 & 23.389 &  0.022 & 21.83 & 1.216 & 0.671 &  3.935 & -3.478 \\
F1 ch1\_1\_s0003 & 210.92911 & 54.367757 & 24.141 &  0.030 & 23.462 &  0.018 & 17.54 & 0.876 & 0.495 &  4.564 & -3.517 \\
F1 ch1\_1\_s0004 & 210.90501 & 54.356026 & 24.407 &  0.035 & 23.465 &  0.023 & 23.93 & 1.097 & 0.672 &  3.608 & -3.459 \\
F1 ch1\_1\_s0006 & 210.91080 & 54.378757 & 24.954 &  0.025 & 23.993 &  0.016 & 17.30 & 0.736 & 0.457 &  4.180 & -3.493 \\
F1 ch1\_1\_s0007 & 210.89823 & 54.376252 & 25.033 &  0.028 & 24.229 &  0.019 & 12.91 & 0.817 & 0.613 &  3.719 & -3.465 \\
F1 ch1\_1\_s0008 & 210.91577 & 54.373818 & 24.554 &  0.024 & 23.490 &  0.012 & 13.55 & 0.691 & 0.341 &  4.226 & -3.496 \\
F1 ch1\_1\_s0009 & 210.90939 & 54.367997 & 24.571 &  0.025 & 23.749 &  0.018 & 14.83 & 0.763 & 0.529 &  3.903 & -3.477 \\
F1 ch1\_1\_s0012 & 210.90896 & 54.369264 & 24.987 &  0.031 & 24.164 &  0.017 & 11.40 & 0.970 & 0.500 &  3.911 & -3.477 \\
F1 ch1\_1\_s0013 & 210.89089 & 54.370105 & 24.981 &  0.027 & 23.841 &  0.018 & 18.75 & 0.891 & 0.557 &  3.335 & -3.442 \\
\enddata
\tablecomments{Measured quantities for 10 Cepheid candidates. A
table listing all 619 candidates is available in the online edition.
Coordinates are in J2000.
}
\end{deluxetable}

\clearpage
\begin{deluxetable}{lllll}
\tabletypesize{\scriptsize}
\tablecolumns{5}
\tablecaption{Wesenheit P-L Relation Slopes in Different Samples\label{Table-2}}
\tablewidth{0pt}
\tablehead{
\colhead{Sample} & \colhead{Slope} & \colhead{Slope$_{err}$} & 
\colhead{[Fe/H]} & \colhead{[Fe/H]$_{err}$}
}
\startdata
Galactic & -3.30 & 0.12 & 0.0 & 0.1 \\
LMC & -3.25 & 0.04 & -0.33 & 0.13 \\
SMC & -3.29 & 0.06 & -0.75 & 0.08 \\
\enddata
\tablecomments{The slope (column 2) and errors (column 3) of the Wesenheit 
period-luminosity relation derived from V and I data in different galaxies 
(as labeled in column 1). Average iron metallicities are 
shown in column 4 with their errors in column 5. There is no readily apparent
dependence of the slope on metallicity. Our M101 slope of
$-3.19 \pm 0.03$ falls within 1-$\sigma$ of the Galactic and LMC values,
and just outside of 1-$\sigma$ for the SMC. 
}
\end{deluxetable}

\end{document}